\documentclass[12pt]{article}
\usepackage{graphicx}
\usepackage{psfig}

\title{\bf The Importance of Few-Nucleon Physics at Low Energy}

\author{H. Arenh\"ovel (1), J. Carbonell (2), L. Canton (3), A. Fonseca (4), \\ W. Gl\"ockle (5), 
H. Hofmann (6), A. Kievsky (7), W. Leidemann (8),\\ G. Orlandini (8), R. Timmermans (9), 
M. Viviani (7)}

\begin{document}

\date{}

\maketitle
\centerline{\it (1) Institut f\"ur Kernphysik, Universit\"at Mainz, Mainz, Germany;}
\centerline{\it (2) Laboratoire de Physique Subatomique et de Cosmologie, Grenoble, France;}
\centerline{\it (3) Istituto Nazionale di Fisica Nucleare, Sezione di Padova, and Dipartimento di Fisica,}
\centerline{\it Universit\`a di Padova, Padova, Italy;}
\centerline{\it (4) Centro Fisica Nuclear, Universidade de Lisboa, Lisboa, Portugal;}
\centerline{\it (5) Institut f\"ur Theoretische Physik II, Ruhr-Universi\"at Bochum, Bochum, Germany;}
\centerline{\it (6) Institut f\"ur Theoretische Physik III, University of Erlangen-N\"urnberg, Erlangen, Germany;}
\centerline{\it (7) Istituto Nazionale di Fisica Nucleare, Sezione di Pisa, and Dipartimento di Fisica,}
 \centerline{\it Universit\`a di Pisa, Pisa, Italy;}
\centerline{\it (8) Dipartimento di Fisica, Universit\`a di Trento, 
and Istituto Nazionale di Fisica Nucleare,}
\centerline{\it  Gruppo Collegato di Trento, Povo (Trento), Italy;}
\centerline{\it (9) Theory Group, KVI, University of Groningen, Groningen, The Netherlands.}

\abstract{This manuscript originates from the discussion at the 
workshop on the
"Future of Few-body Low Energy Experimental Physics" (FFLEEP),
which was held at the University of Trento on December 4-7, 2002
and has been written in its present form on March 19, 2003.
It illustrates a selection of theoretical advancements in
the nuclear few-body problem, including two- and many-nucleon
interactions, the three-nucleon bound and scattering
system, the four-body problem, the  A-body (A$>$4) problem,
and fields of related interest, such as reactions of
astrophysical relevance and few-neutron systems.
Particular attention is called to the contradictory situation
one experiences in this field: while theory is
currently advancing and has the potential to inspire
new experiments, the experimental activity is
nevertheless rapidly phasing out. If such a trend
will continue, advancements in this area will become
critically difficult.}

\vfill\eject

\tableofcontents



\section{Introduction}

In nuclear physics one studies interacting hadrons in the nonperturbative
regime of quantum cromodynamics (QCD). This field faces two natural frontiers. The first one is 
devoted to the origin and the fundaments of the strong force between the 
nucleons, as well as its parametrization in terms of effective degrees of 
freedom (d.o.f.), suitable for the nonperturbative regime. The other is 
the completely microscopic, quantum mechanical treatment of many-body 
systems (nuclei) in terms of these effective d.o.f.\ with the strong 
force as dynamical input. 
Within these two frontiers, few-nucleon 
physics at low energy represents quite a rich and flourishing research field 
describing a great variety of strong interaction phenomena. 
%
%
The particular role of this field for the whole of nuclear physics 
is illustrated and underlined by the following three statements:

\begin{itemize}

\item ``Why few nucleons?'' The relatively small number of particles allows 
accurate solutions of the quantum mechanical many-body problem without 
the need of approximations, necessary and unavoidable for more complex 
systems. Therefore 
the comparison of such theoretical results with experimental data of 
comparable accuracy can lead to conclusive statements with respect to 
the assumptions for the underlying nuclear dynamics on which the theory 
is based.

\item ``Why at low energy?'' The main interest is to understand strong 
hadron dynamics in the {\it nonperturbative} regime in a well controlled 
{\it nonrelativistic} framework. Therefore observables at {\it low} 
energy and momentum transfers represent the best testing grounds. 
Furthermore, possible {\it relativistic effects} can be handled quite reliably
by incorporating leading order relativistic contributions consistently 
without the need of a completely covariant approach.

\item ``Why could it be flourishing?'' It is a field of ``low cost 
investments'' with ``high revenues'' in terms of fundamental insight. 
The theoretical methods are accurate and under control, and 
experiments are considerably much less costly than is the case 
for high-energy observables.
\end{itemize}

In this note we would like to point out the major achievements of 
few-nucleon physics at low energy with respect to the two above mentioned 
frontiers, and what are the open problems at present. But this note 
is not intended to give a comprehensive review of the impressive 
progress which has been obtained in recent years in this field, rather a 
few, particularly illustrative examples have been selected in order to 
underline the punch line of this note, namely 
{\it to prevent a complete shut-down of low-energy machines} 
for the study of few-nucleon systems and 
to fuel a {\it more substantial support} of experiments using such 
machines. Besides illustrating the progress, the selected examples will 
point out important questions and open problems of fundamental nature that 
have to be addressed in the future and for which a continuation of ongoing 
experiments is mandatory at appropriate existing accelerator facilities as 
well as at new dedicated ones.

The note is organized as follows. Sec.~\ref{2_body_syst} is devoted
to the nature of the strong interaction and its underlying degrees of freedom. 
Present models of the nucleon-nucleon (NN) potential 
will briefly be reviewed as 
obtained by studying bound and scattering states of the two-body system. 
Electromagnetic interactions with 
the two-nucleon system provide further insight into the properties of those 
forces and will be discussed in this section, too. In particular, we will 
suggest possible future experiments needed for a more detailed 
analysis of the underlying hadron dynamics. Sec.~\ref{many_body_interaction} 
is devoted to three- and four-body systems with 
special emphasis on the three-nucleon force (3N-force). 
The specific role of three-body nuclei for analyzing models of the 
3N-interaction is discussed in detail. Also the four-body system is 
still "light" enough allowing 
accurate microscopic calculations in order to shed additional
light on the 3N-force as well as on possible manifestations of 
four-nucleon forces (4N-forces). 
Furthermore, its characteristics of large binding energy, large central 
density and ``closed-shell'' structure closely resembles heavier systems. 
Therefore, it can be considered as an instructive testing ground for many-body 
approaches, models and approximations. 
The accurate treatment of many-body systems larger than four will be 
addressed in Sec.~\ref{many_body_systems}. 
It will be pointed out how few-nucleon 
physics can contribute to understand the dynamics of more complex nuclei 
on a microscopic basis. In particular, the phenomena of clusterization and 
collective motion will be discussed. A few examples will be shown, where 
{\it ab initio} microscopic calculations of bound states and reactions  
in the continuum in ($A>4$)-systems start showing typical 
cluster or collective features. Again the need of appropriate 
experiments to clarify these issues will be emphasized. 
In Sec.~\ref{few_body_slave}, the important role of few-body physics for
other fields, like elementary particle physics and astrophysics, is 
described very briefly by pointing out the use of very light nuclei 
as effective neutron targets and giving a few examples of few-body reactions
of astrophysical relevance. Finally, concluding remarks are given in 
Sec.~\ref{conlusions}.



\section{Study of the Hadronic Force in the Two-Body System}
\label{2_body_syst}
The first step for developing a model for the strong nucleon-nucleon (NN) 
interaction is
to study the two-nucleon system, its bound and scattering states. 
The bound state properties are well known and nowadays a wealth of 
NN-scattering data exist, accumulated over many years. These data
serve as input for fitting models of the NN-interaction. In the early
days of nuclear physics very simple and purely phenomenological models 
have been introduced. After the basic idea of Yukawa of mesons as 
mediators of the strong interaction more sophisticated models have
been developed. In fact, for many years two-nucleon physics has dealt 
with the design of more and more refined NN-potentials, 
parametrizing the NN-force in terms of meson and 
nucleon degrees of freedom, which, with the advent of QCD, 
are now considered as effective only. 

In order to construct such NN-potentials one needs as a first
step a partial-wave analysis (PWA) of the scattering data. In
recent years, new methods of energy-dependent PWA have made
possible the accurate determination of the NN phase-shift and
mixing parameters. In order to achieve a high-quality description
of the presently available database, a sophisticated treatment
of the long-range electromagnetic interaction is required. Also
the one-pion exchange interaction needs to be included, as well
as a model for the medium-range interaction (heavy-meson exchange
or two-pion exchange).

At present, a variety of so-called high-precision 
NN-potentials is available which fit the 
NN-scattering data (available at the time of construction)
with a $\chi^2$ per datum of about one. 
Thus they describe those NN-data with magnificent accuracy. 
Beyond the longest range one-pion-exchange (OPE) part, 
which all these potentials contain, the 
medium and short range region is parametrized either purely phenomenologically 
or semiphenomenologically by the exchange of heavier mesons and introduction 
of hadronic vertex form factors. These modern conventional NN-forces 
contain typically 40-50 parameters but describe NN scattering data with high 
precision up to 350 MeV laboratory energy. With respect to the short range
region, also more microscopic models with quark and gluon d.o.f. have 
been introduced. 

Besides these 
phenomenological and meson-theory based models, recently another approach 
has been put forward in the framework of effective field theory (EFT) 
using only nucleon and pion d.o.f. It is based on chiral 
perturbation theory and starts from the most general 
Lagrangean for pion and nucleon fields obeying spontaneously broken chiral 
symmetry of QCD. The nuclear forces are constructed via a systematic expansion 
with respect to a small scale parameter, which is given by the ratio of 
generic external momenta $Q$ and a mass scale $\Lambda$ of the order of 
the $\rho$-meson mass. 
For momenta below $\Lambda$, nuclear forces can be evaluated 
systematically in increasing order of that ratio at the cost of an 
increasing number of contact forces. In leading order (LO) there is 
the well established one-pion exchange supplemented by contact forces which 
subsume unknown short range physics and which are fixed by fitting 
low-energy observables. The strength factors of those contact 
forces are called low-energy constants (LEC's). In the next to leading order 
(NLO) there are various two-pion exchange diagrams and additional contact 
forces with new unknown LEC's. Even higher order forces with more LEC's 
have been already worked out.

Up to now chiral NN-forces have been fitted to NN-scattering data within NNLO,
which amounts to fixing nine LEC's. A good description of the standard and well
established NN-phase shift parameters has been achieved up to about 100 MeV. 
In contrast to conventional high-precision NN-forces, 
which are not based on a systematic expansion, the fit is not perfect. 
In fact it makes no sense to expect high-quality fits at low chiral order. 
More precise fits are expected for higher order at the expense of more
LEC's. It remains to be seen up to which order one has to go for 
reaching the same precision and whether the number of LEC's 
will then be comparable to the typically 40-50 
parameters of the most modern conventional NN-forces. 
At present an interesting and lively debate between the traditional meson 
theory point of view and the EFT approach is going on and will presumably 
continue in the years to come where one important question is related to the 
model dependence or model error. 

With respect to the experimental scattering data we would like to remark
that the presently available database for $pp$ scattering is
of good quality. However, for $np$ scattering the situation is not
quite as good. New experiments, both $pp$ and $np$, should be aimed
at improving the existing phase shifts. What new types of data are
still useful is a question that can be answered in a quantitative
manner by a collaborative effort of theorists and experimentalists,
by using the existing partial-wave analysis as a benchmark.

Having determined the NN-force, one can proceed to study the theoretical 
predictions on the hadronic and electroweak structure of light nuclei. 
A central question is: how well do we understand those properties which 
were not used in fixing the parameters of the NN-interaction? In this sense,
few-body nuclei serve as test laboratories for the study of a variety of 
hadronic and electroweak properties. A particularly well suited tool is 
provided by the electromagnetic (e.m.) probe, and in fact e.m.\ reactions 
on the two-nucleon system have allowed further insight into the 
structure of the hadronic force. The reason for this is that the e.m.\ 
interaction proceeds via the e.m.\ current of the hadronic system 
which requires the explicit knowledge of the d.o.f.\ underlying the 
interaction. In fact, the e.m.\ current has to be consistent with the 
Hamiltonian, because gauge invariance alone is not sufficient for its full 
determination. 

For the two-body sector quite 
elaborate theoretical results are available describing a variety of 
e.m.\ processes on the deuteron like, e.g., static moments, elastic 
electron scattering, photo- and electrodisintegration and meson production.
Present most advanced theories include consistent two-body meson exchange 
currents (MEC), 
isobar configurations and relativistic contributions of leading order in
a $p/M_N$ expansion. 

At low energy and momentum transfers, gross properties like, e.g., 
total photoabsorption cross section, 
elastic form factors and inclusive response functions
are in satisfactory agreement with experimental data on the 5 percent
level, and within this accuracy one finds almost no sensitivity to the
theoretical input, i.e., to the realistic NN-interaction model and 
corresponding induced currents. For these gross properties, differences 
with respect to different interaction models appear on a much smaller scale, 
namely on the one percent level, requiring experimental data of at least 
the same accuracy which unfortunately do not exist. 
For deuteron photodisintegration a critical assessment of existing 
experimental data on various observables and a detailed comparison to
the theory was given about ten years ago. 
Thus for a more detailed comparison between experiment and theory one
desperately needs new experimental data of considerably higher accuracy. 
This is particularly true for more exclusive observables like differential
cross sections and polarization observables which show in certain kinematic
regions a considerably higher sensitivity on the level of 10-20~\% 
to the theoretical input. In particular, 
polarization observables
bring often a much richer insight and stimulate in addition experimental 
interest due to technical challenges. Of special interest is also the 
transition between nonrelativistic and relativistic regime which appears 
in some observables already at quite low energies.

A few examples might be useful in order to illustrate the urgent need of
precise experimental data:
\begin{enumerate}
\item[(i)]
Total photodisintegration cross section and spin asymmetry. The latter 
determines the much disputed Gerasimov-Drell-Hearn sum rule and in fact,
a very large negative contribution is expected very close to the break-up
threshold because of the dominant magnetic transition to the $^1S_0$ scattering
state which contributes only when photon and deuteron spins are opposite. 
In particular, a sizeable influence of relativistic contributions has been 
found in the spin asymmetry at low energies, say a few MeV above break-up
threshold.
\item[(ii)]
Nucleon polarization in $d(\gamma,N)N$ at low energy. There exist
very few data which are of rather low accuracy. Also very few data on 
vector and tensor target asymmetries exist. 
\item[(iii)]
Exclusive electrodisintegration at low energy and momentum transfer 
$d(e,e'p)n$. Recent experimental results indicate a problem
in the longitudinal-trans\-verse interference structure function $f_{LT}$
which seem to be present also in the transverse one $f_T$. The latter, 
however, could not be separated from the longitudinal one in that experiment. 
It requires a Rosenbluth separation. 
\end{enumerate}

Finally, we would like to point out that NN bremsstrahlung is
another very interesting electromagnetic reaction in the two-body sector, 
which only recently, with the advent of high-precision data for
$pp$ bremsstrahlung, has become a good testing-ground
for NN-interaction models. At present, theoretical models fail
to describe the available $pp\gamma$ cross sections. The size of
the discrepancy between theory and experiment is surprisingly
large.




\section{Many-Body-Forces in Three- and Four-Body Systems}

\label{many_body_interaction}

An important new feature appears in few-nucleon systems with more than
two particles: in the framework of presently known force models, these
systems cannot be described using NN-forces only. It is necessary to
include three-nucleon forces (3N-force). The origin  and the explicit
form of the 3N-force will be a central issue of future  few-nucleon
physics. 
 
While the development of realistic NN-force models was possible thanks 
to a large amount of experimental NN-scattering data at a wide 
range of energies, the study of the 3N-force relies at present
on a very small data basis only (binding energies, low-lying excited 
states of light nuclei, and a few reactions). It is clear that for a 
reliable and realistic 3N-force model, based on a deeper understanding 
of the underlying dynamics, the experimental data set has to be enlarged
considerably if one wants to reach the same degree of accuracy as in the
NN case. 

It is important to note that 3N-forces are not introduced in an ad hoc
manner. Meson exchange theory as well as EFT and a more fundamental view
in terms of quark-gluon dynamics naturally suggest that additional
many-body forces between nucleons exist  which cannot be reduced to
pair-wise NN-interactions. Three-nucleon forces were proposed already
in the thirties.  Much later, first quantitative studies
were carried out,  where an important 3N-force contribution, arising
from the  intermediate excitation of a nucleon into a $\Delta$ in a
two-$\pi$-exchange,  was included in the three-nucleon ground 
state. It became clear that such effects cannot
be neglected for the three-nucleon binding. 

Without theoretical guidance, however, the phenomenological approach
towards  3N-force is rather hopeless. In contrast to NN-forces, where
only a few basic operators can contribute, the number of possible
three-body operators is prohibitive, leading to a tremendously large 
number of terms, which moreover are accompanied by unknown scalar
functions  depending on four relative momenta. 

Although the natural testing ground for the 3N-force is the
three-nucleon system it turns out that some scattering observables
over a broad range of energy show little sensitivity to the 
NN-force model choice or the inclusion of the 3N-force. This creates
additional requirements for precision data needed to validate ongoing
theoretical research.  As will be shown below four-nucleon forces seem
to be very small and thus also the four-nucleon system can play an
important role in the investigation of the 3N-force. In addition, as
will be pointed out in  Sec.~\ref{many_body_systems}, important
additional information on the 3N-force comes from the study of the
low-lying spectra of light nuclei with A~$>$~4.
 
\subsection{Bound States}\label{bound_states} 

Nowadays three- and four-nucleon bound states and binding energies can
be calculated with different methods based on any of the most modern 
high-precision NN-forces with an accuracy on the percentage level or 
less. For A=3 and 4 a well founded formulation, 
the Faddeev-Yakubovsky scheme (FY),
opened that avenue followed by alternative, equally accurate procedures:
expansions in hyperspherical harmonics (HH) 
or gaussians (CRCGV), stochastic variational method
(SVM) and path integral techniques in form of the 
``Green's Function Monte Carlo'' method (GFMC).
Recently other very promising methods, based on the theory of effective 
interactions, have been developed: the ``no-core shell model'' 
(NCSM) and the ``effective interaction HH''
(EIHH) using expansions in harmonic oscillator and HH basis
functions, respectively. 
 
\begin{table}[tbp] 
 \caption{Kinetic $\langle T \rangle$, potential $\langle V \rangle$,
binding energy $E_b$ (all in MeV),
and mean square radius of $^4{\rm He}$ as obtained by various 
methods.}
\begin{center}
\begin{tabular}{lllll}
  \hline
Method& { $\langle T \rangle $ } & {$\langle V \rangle $} & { $E_b$  } & 
{ $\langle r^2\rangle^{1/2}$}\\
  \hline
FY     & 102.39     & -128.33      & -25.94(5)   & 1.485     \\
CRCGV  & 102.25     & -128.13      & -25.89      &          \\
SVM    & 102.35     & -128.27      & -25.92      & 1.486     \\
HH     & 102.44     & -128.34      & -25.90(1)   & 1.483     \\
GFMC   & 102.3(10)  & -128.25(10)  & -25.93(2)   & 1.490(5)  \\
NCSM   & 103.35     & -129.45      & -25.80(20)  &           \\
EIHH   & 100.8(9)   & -126.7(9)    & -25.944(10) & 1.486 \\
  \hline
 \end{tabular}
\label{table1}
\end{center}
\end{table}

\begin{table}[h]
\caption{Binding energies of light nuclei in MeV based on the
NN-potentials CD~Bonn, Nijmegen~I and
AV18.}
\begin{center}
\begin{tabular}{ l r r r r }
\hline
       & CD-Bonn & Nijm I & AV18 & Exp  \\
\hline
$^3$H  &  8.01 &  7.74 &  7.6 &  8.48 \\
$^4$He & 26.26 & 24.98 & 24.1 & 28.30 \\
$^6$He &       &       & 23.9 & 29.27 \\
$^6$Li &       &       & 26.9 & 31.99 \\
$^7$He &       &       & 21.2 & 28.82 \\
$^7$Li &       &       & 31.6 & 39.24 \\
$^8$He &       &       & 21.6 & 31.41 \\
$^8$Li &       &       & 31.8 & 41.28 \\
$^8$Be &       &       & 45.6 & 56.50 \\
\hline
\end{tabular}
\label{table2}
\end{center}
\end{table}

An example of the degree of accuracy reached by these methods in the
four-body  system is given in Table~\ref{table1}, where 
binding energy, expectation values of kinetic and potential energies and 
mean square radius of $^4$He are shown as they result from calculations
of seven different  techniques based on the same potential model.
These most precise techniques lead to a clear cut answer: all modern
realistic  NN-forces underbind light nuclei significantly. Examples are
displayed in Table~\ref{table2}. The results can be considered as
evidence for the  existence of at least a 3N-force.

\begin{table}[tbp]
\caption{Theoretical binding energies for $^3$H and $^4$He in MeV 
based on the 3N-forces TM' and Urbana IX.}
\begin{center}
\begin{tabular}{ l r r r r }
\hline
                & $^3$H & $^4$He  \\
\hline
CD-Bonn + TM'    &  8.48 & 28.4   \\
AV18 + TM'      &  8.45 & 28.36 \\
AV18 + Urbana IX &  8.48 & 28.50 \\
Exp. &  8.48 & 28.30 \\
\hline
\end{tabular}
\label{table3}
\end{center}
\end{table}

As already stated above, the origin and the nature of 3N-forces as 
well as their description, consistent with the NN-force, is one of the
central  questions in the years to come. At present 3N-force
models based on two-pion-exchange mechanism are 
available. Several of them, e.g., Tucson-Melbourne (TM) and 
Urbana IX have been employed in the above mentioned 
calculations and their parameters have been  adjusted to the $^3$H
binding energy. It is interesting to notice that  the inclusion of
3N-force leads to a nearly correct 
$\alpha$-particle binding energy as we see in Table~\ref{table3}. 
There is a small overbinding which can presumably be removed by an
additional  fine tuning of the 3N-force parameters. In any case, in view
of the high central density of the $\alpha$-particle, these results
strongly indicate that four-nucleon forces presumably are
small.

Besides the extensively investigated 3N-force based on
two-pion-exchange mechanisms, interesting additional structures for 
the 3N-force appear for the exchange of heavier mesons. One finds a
variety  of short-range contributions including structures
that are  quite similar to those recently discussed also in the
framework of  EFT. Additional 3N-force terms appear if
one combines  pion-emission and re-absorption processes  with the
dynamics of the 3N system. 

 \begin{figure}[tbp]
   \centering
\includegraphics[scale=0.5]{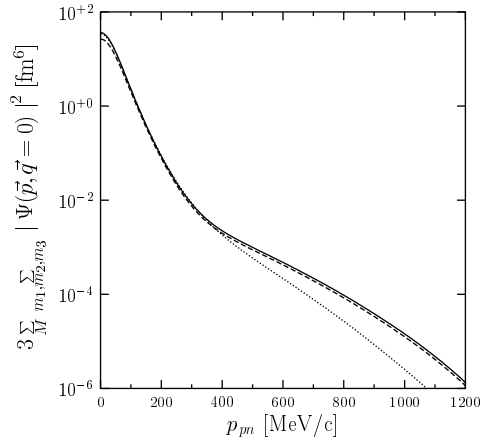} 
  \caption{The two-nucleon relative momentum distribution in $^3$He. The
vectors
 $\vec p$ and $\vec q$ denote standard Jacobi momenta. Solid curve
 stands for  AV18, the dashed curve for AV18 +Urbana IX and the dotted
 curve for CD-Bonn.}
   \label{n2_p}  
  \end{figure}  
 
 \begin{figure}[tbp]
  \centering
   \includegraphics[width=0.5\textwidth]{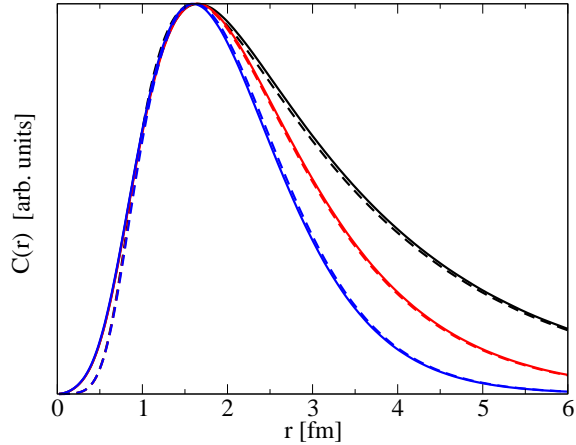}
  \caption{Two-nucleon correlation functions for light nuclei. The paired
curves from below to above beyond the peak are for the $\alpha$-particle,
for $^3$He, and for the 
deuteron, respectively. Solid curves represent
results obtained with CD~Bonn+TM (CD~Bonn for the deuteron) and dashed
curves with  AV18+TM (AV18 for the deuteron).}
 \label{C_r}
 \end{figure}  
 
As an illustration of the status of the theory revealing
properties of bound state wave functions, we show in
Fig.~\ref{n2_p} the two-nucleon relative momentum distribution
for $^3$He, and in Fig.~\ref{C_r} the two-nucleon correlation functions
$C(r)$ for light nuclei. In Fig.~\ref{n2_p} we see different 
populations of relative
momentum distribution for CD~Bonn and AV18. The
3N-force effects are rather small in all cases. Interestingly,
the probabilities to find two nucleons at a  distance
$r$ in the deuteron,
$^3$He, and $^4$He are essentially identical for $ r \le  2$ fm and
deviate at larger distances only due to different binding energies
(Fig.~\ref{C_r}).

Although these wave function properties help us having a 
"pictorial" representation of the dynamical effects, they are not
directly  observable; nevertheless, they may leave their mark on certain
experimental observables, allowing thus their study indirectly. To this
end it is necessary to have  a complete control on all other theoretical
ingredients relevant to the considered observable so that the
dependence on the remaining  quantity may be studied in a reliable
manner.  The determination of the neutron electric form factor from
electron  scattering off $^2$H and $^3$He by studying polarization
observables  is a typical example for this (see
Sec.~\ref{neutron_target}). 

\subsubsection{Neutron-Neutron Interaction and Neutron Clusters}

A recent experimental result suggests the possible detection
of bound multi-neutron clusters (A~$\ge3$). Such a possibility, in
particular concerning a tetra-neutron, has been recurrently considered in
the literature both from the experimental and theoretical side but never
confirmed.

The experimental method is based on beams of
neutron-rich nuclei and differs from the previous ones
in that the $4n$-formation is no longer based on double charge exchange
processes (e.g. $\pi^-$ + $^4$He $\to \pi^+$ + $^4n$) but rather in
removing from already pre-existing $4n$-clusters the parasites which impede
it to exist. Thus, studying the break-up  of $^{14}$Be by a carbon
target, some events compatible with the reaction $^{14}$Be $\to ^{10}$Be
+ $^4n$ were reported. The number of these events, though well separated
from the background,  is however very small. New runs with increasing
statistics are necessary to confirm  or disprove the findings.

From  the theoretical side it seems clear that the $nn$-force alone
cannot bind such a system. Despite a quasi-resonant $nn$-scattering 
length ($a_{nn}\approx-18$~fm), the Pauli principle generates
a strong effective repulsion which prevents a few-neutron cluster to
exist. The required additional attraction should eventually come from
$3n$-forces only. In recent GFMC calculations a variety of 
$3n$-interactions
have been explored for fitting the A~=~8 spectrum, and the authors were
very pessimistic about  the existence of small
neutron droplets. At the same time, they point out our poor
knowledge of the $3n$-force. The Urbana TNI-model, for instance,  which so
successfully describes the 3N- and 4N-phenomenology, turned out to fail
when extrapolated to neutron-rich nuclei. The different versions of the
Illinois forces, built to overcome these difficulties, give binding
energies which vary over a range of about 1.5~MeV for a $T=2$ state like
$^8$He. 
In the $4n$-system, the missing  attraction is far beyond such 
uncertainties and its existence is very unlikely. The situation 
is less clear for $A>4$. 

It is certainly a very interesting and  
challenging task to draw a definite conclusion on the existence of small
neutron clusters given the non-existence of a di-neutron and theoretical
results advocating unbound neutron matter.

\subsection{Scattering States}\label{scattering_states}

Bound state properties provide only limited information on the nature
of the hadronic force in nuclei, whereas scattering states, appearing 
in reactions with hadronic and electromagnetic probes, yield a much 
richer and much more detailed information. A particularly rich source 
is obtained by studying polarization observables.

\subsubsection{Three-Body Continuum}\label{3N_system}

In contrast to the two-body case, continuum states for three particles
are a non trivial theoretical problem. However, several powerful tools
have been developed in the past in order to calculate scattering states
in the continuum using realistic potentials: the  Faddeev
approach or the reformulation by the AGS scheme
as well as the Kohn-variational approach using hyperspherical
expansions. 

 \begin{figure}[tbp]
  \centering
  \includegraphics[width=.9\textwidth]{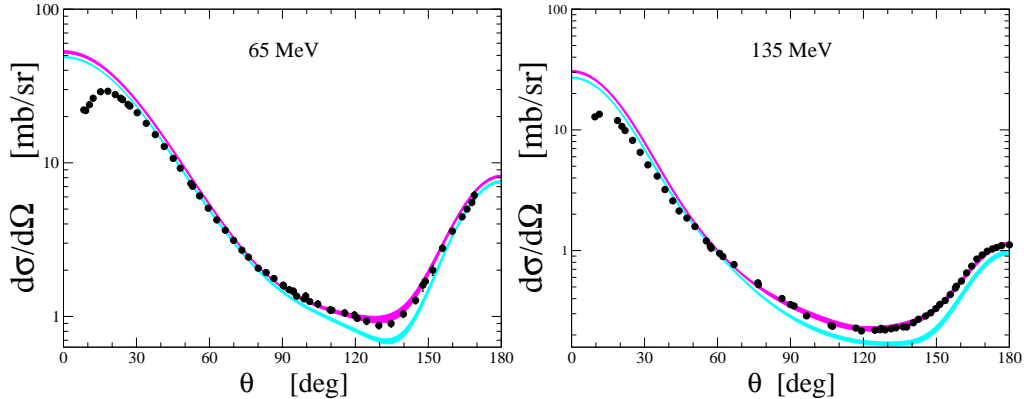}
  \caption{Angular distribution in elastic $nd$ scattering  
(light grey band: modern NN-force predictions only; dark grey band:
including 3N-forces) compared to $pd$ data}
  \label{nd_el}
 \end{figure}
 
\begin{figure}[tbp]
   \centering
\includegraphics[width=.5\textwidth]{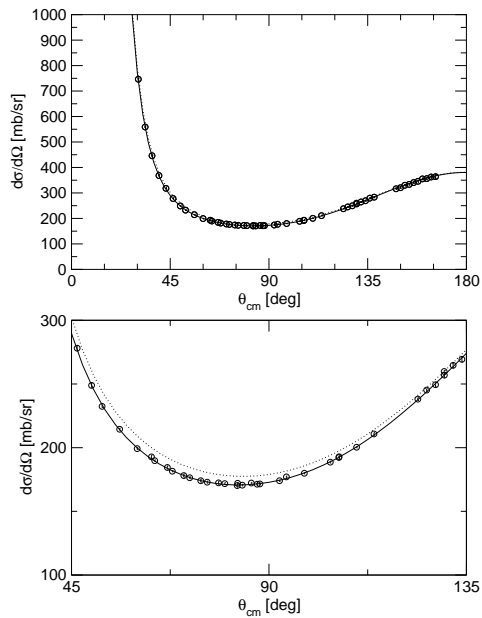}
  \caption{Differential cross section versus $\theta_{CM}$ for $p+d$
elastic scattering at $E_p~=~1$~MeV.}
   \label{pisa-EP1MeV} 
  \end{figure}

This progress has opened up the possibility to access the tremendous 
variety of spin observables and multi-nucleon fragmentations in hadronic 
as well as electromagnetic reactions. This will be illustrated by a few
cases.  Fig.~\ref{nd_el} displays the angular distributions of 
elastic  $nd$  scattering at 65 and 135 MeV using NN-forces alone,
and showing the effect of adding 3N-forces like the ones in 
Table~\ref{table2}. One readily notes a clear cut discrepancy 
in the minimum of the data with NN-force predictions only, and 
a nice agreement when 3N-forces are added (parameters of the 3N-force
have been fixed before on the $^3$H  binding energy). As to the discrepancy 
at forward angles one has to notice that the experimental data refer
to  $pd$  scattering and the Coulomb force should be included in the theory.
Although this result is very promising, much work remains to be done.
While some observables in elastic $nd$ scattering can be well
described by NN-forces alone and addition of  3N-forces plays no
significant effect, others are only well described if 3N-forces are
included. Some of these effects can even be seen at low
energy, as shown in Fig. \ref{pisa-EP1MeV} for $pd$ elastic scattering at
1 MeV. The Coulomb force between protons is included in a very precise
calculation that uses a correlated hyperspherical expansion of the wave
function in configuration space together with the complex Kohn
variational principle to determine the $S$- or the $K$-matrix. Part of
this effect may be understood in terms of scaling with $^3$He binding
energy when the 3N-force is included but clearly demonstrates the need
for very precise data. Nevertheless one finds that for specific spin
observables there are striking discrepancies to the data with present
force models.

 \begin{figure}[tbp]
  \centering
  \includegraphics[width=.5\textwidth]{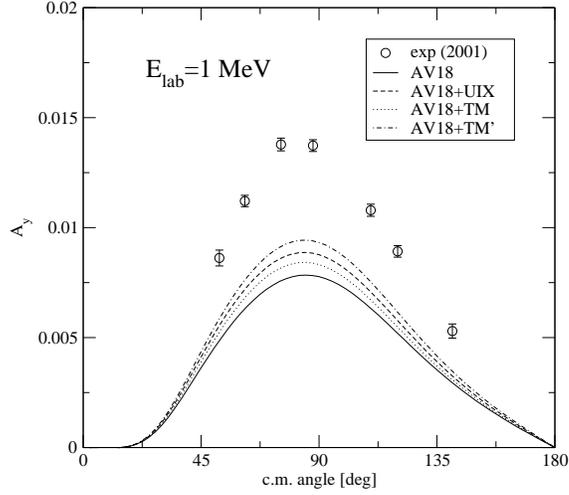}
  \caption{Proton analyzing power $A_y$ for elastic $pd$ scattering
showing the $A_y$-puzzle; curves: theoretical results using various
interaction models (see inset).}
  \label{figAy}
 \end{figure}

There is a well-known discrepancy in the description of the vector
analyzing powers $A_y$ (``$A_y$-puzzle'') and $iT_{11}$ for $Nd$
scattering. The comparison of theoretical predictions with data at 
$E_{lab}\le 2$~MeV shows that the calculations
underpredict  the data by $\approx$ 30~\% in the maximum of the angular
distributions. The disagreement becomes worse at lower energies  where
the difference increases to $\approx$ 40 \% at
$E_{lab}=650$ keV. None of the modern
NN-potentials is able to  describe $A_y$. Even when 3N-forces such as TM
or Urbana IX are added, the  discrepancy is only slightly reduced (see
Fig.~\ref{figAy}).  The discrepancy in $A_y$ and $iT_{11}$ decreases
with increasing energy.  All this clearly indicates that
the spin-isospin structure of the 3N-force  is not yet sufficiently
well understood and that much more experimental and theoretical work is
required. New theoretical ways are already being  explored, where purely
phenomenological, meson theoretical, or EFT
approaches are used.  

 \begin{figure}[tbp]
\centerline{\includegraphics[angle=-90,width=.495\textwidth]{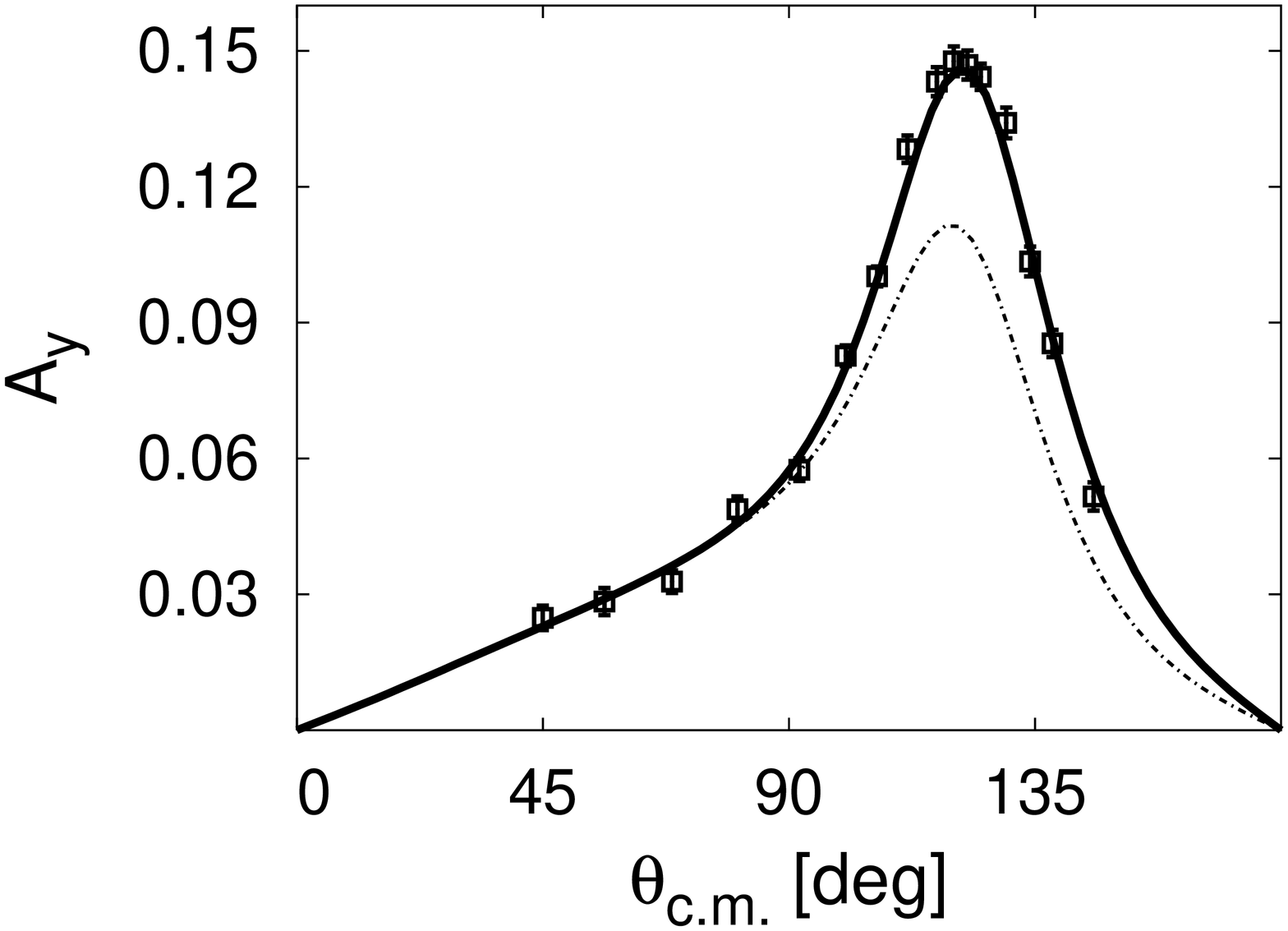}}
  \includegraphics[angle=-90,width=.495\textwidth]{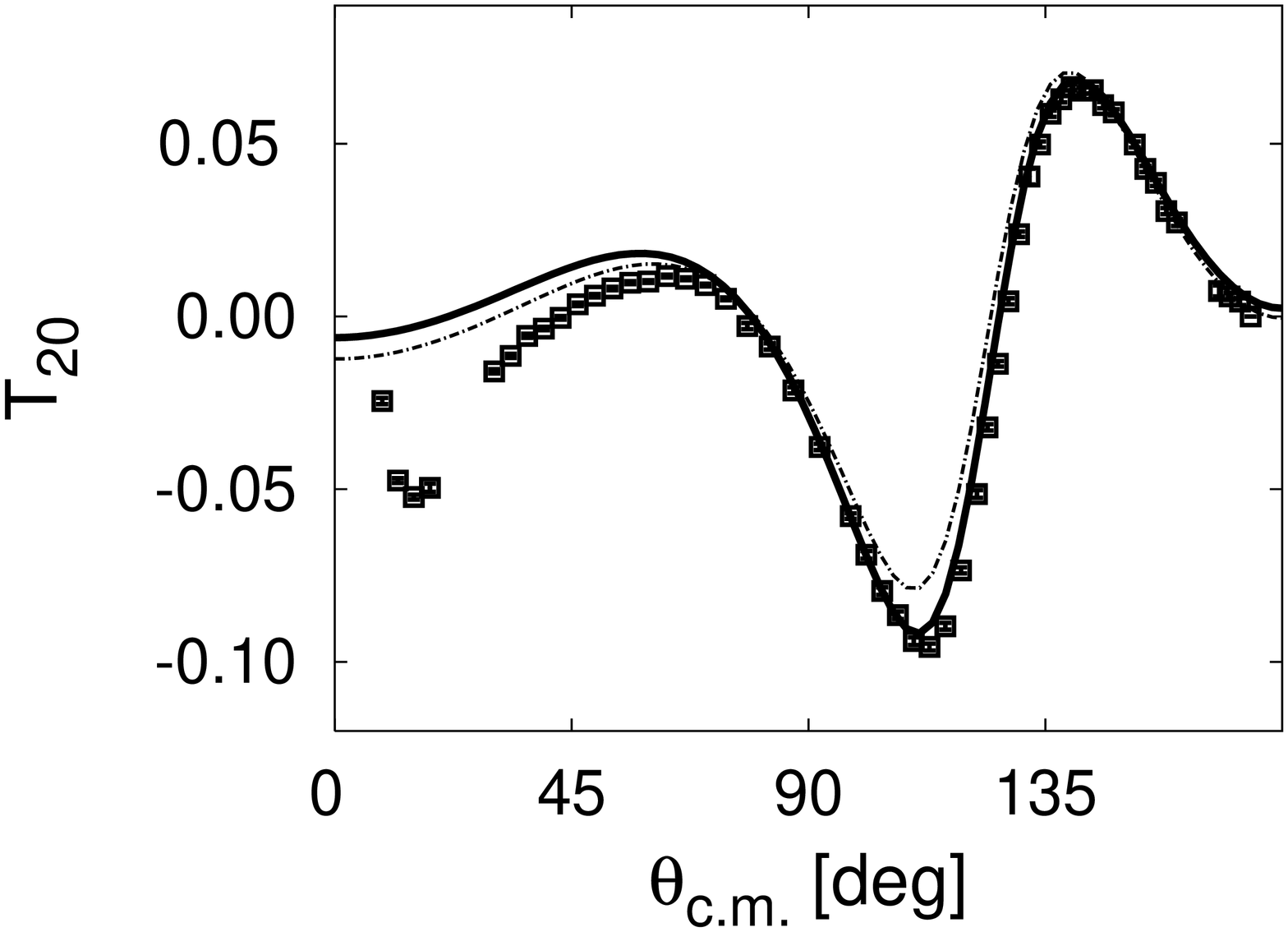}
  \includegraphics[angle=-90,width=.495\textwidth]{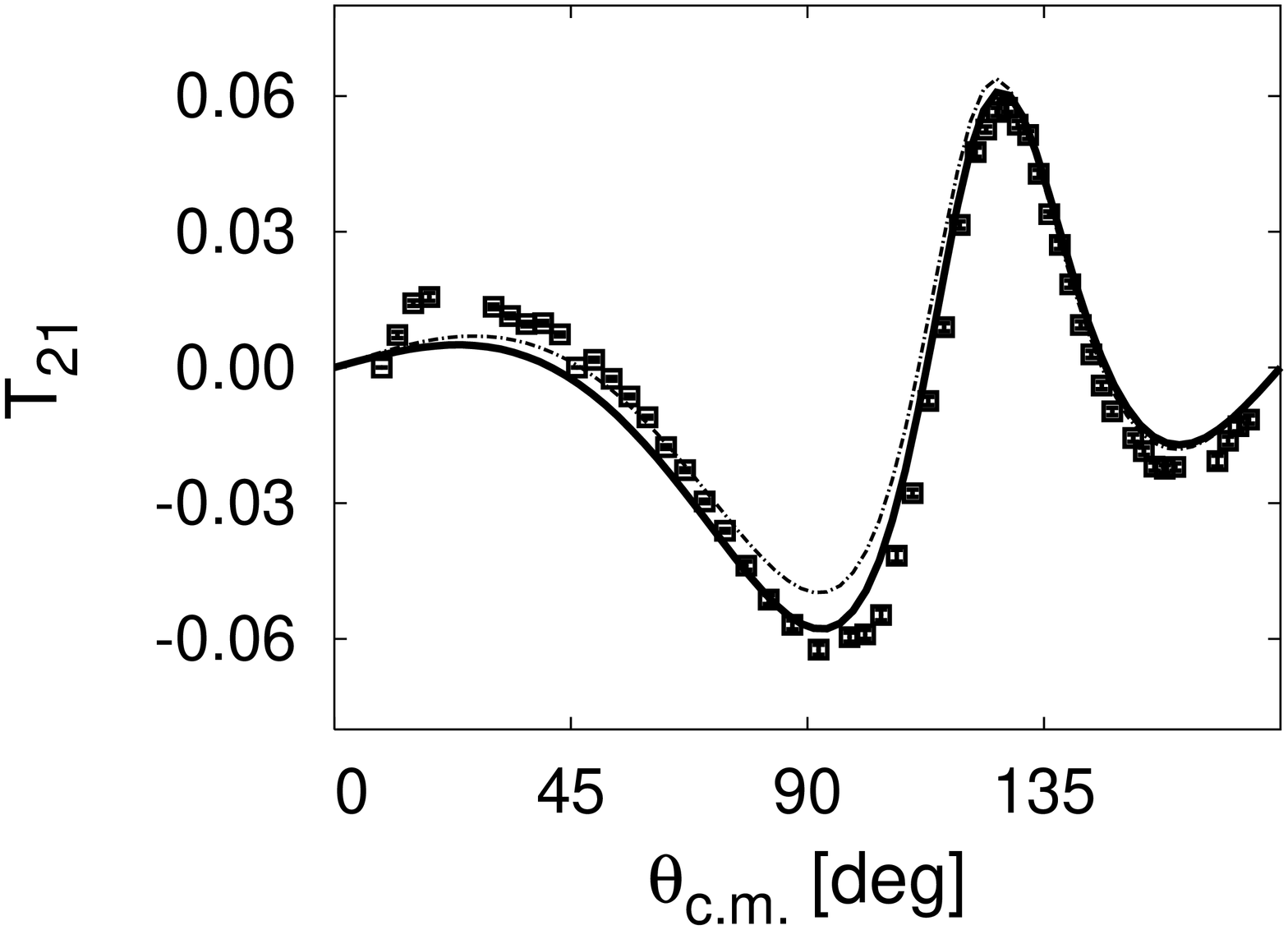}
  \caption{Spin observables for $nd$ scattering at 8.5 MeV. Solid
(dash-dot) line including (excluding) the
3N-force. For $T_{20}$ and $T_{21}$ comparison is
made with $pd$ data measured at energies shifted by $+0.7$ MeV.}
  \label{Ay}
 \end{figure}

 In a recent meson theoretical calculation, pion retardation effects are
taken into account. That such retardation effects can
contribute to the  3N-force has been known since at
least for two decades. In this scheme, single-pion exchanges lead to a
non-negligible 3N-force of tensor  structure, similar to the OPE term
of the NN-interaction  but involving the coordinates of all three
nucleons. It leads to a distinct  energy dependence of the 3N-force. At
the present stage, there are still theoretical uncertainties which
require the introduction of one adjustable parameter, but for the
future one hopes that this mechanism may be constrained
theoretically in order to obtain a completely  parameter-free
description of this effect. With a suitable 
adjustment of this single parameter one can remove the discrepancies 
for the vector analyzing powers, with smaller (but significant) 
effects for the other spin observables, such as the tensor analyzing
powers  (see Fig.~\ref{Ay}). This is another example on how 
additional precise measurements can help to clarify the role of new 
pion-exchange terms in the three-nucleon dynamics, and consequently, 
can lead to a deeper understanding of the role played by the pion 
itself in the three-nucleon system.

While suffering from a certain lack of systematics in the organization
of the dominant contributions, the meson picture has the
advantage of visualizing the process by means of a particle exchange
diagram, and to interpret intuitively the origin of the various terms.
It also provides a direct link between the behavior of the few-nucleon
systems at low energies and nuclear phenomena at intermediate energies,
above the meson-production threshold where retardation is
indispensable. This is an important aspect: 
few-nucleon physics at low-energy does not represent an isolated
research field, but is strongly connected to contiguous areas like
hadronic physics at intermediate  energies and many-body physics, with
vast opportunities  for cross-fertilization.

 \begin{figure}[htb]
  \centering
  \includegraphics[scale=0.5]{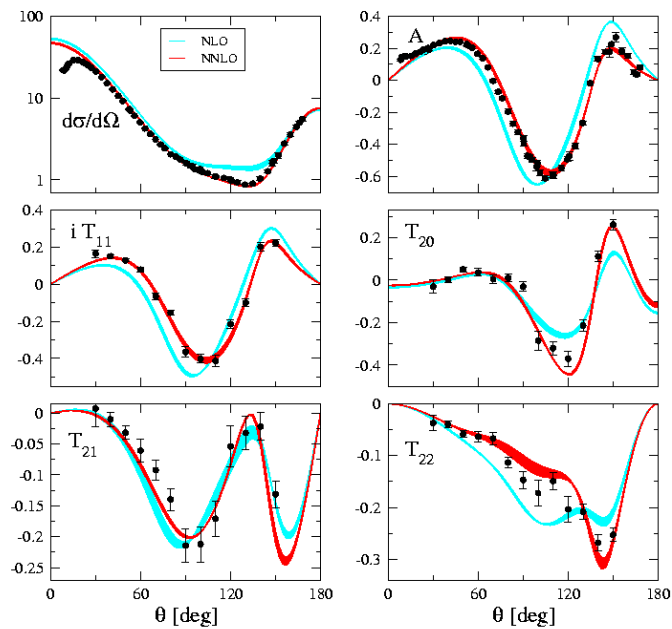}
  \caption{Elastic $nd$ scattering at 65 MeV in NLO (grey band) and NNLO 
(dark band) of EFT approach compared to $pd$ scattering data}
  \label{elast2_65mev}
 \end{figure}

The chiral approach on the other hand has the advantage of a consistent
simultaneous determination of NN- and 3N-forces. In this respect, the
guidance  provided by chiral perturbation theory can be very useful. In
the chiral approach the 3N-force appears at NNLO  and depends on two
new parameters. This 3N-force appears in three topologies: a two-pion
exchange, a one pion exchange between a two-nucleon contact force and
the third nucleon and a pure 3N-contact force. The two new parameters
are related to the latter two pieces , whereas the two-pion exchange is
parameter free.  The two parameters are adjusted to the $^3$H binding
energy and the $nd$ doublet scattering length.  Although this
approach cannot at this time provide a cure for the $A_y$
discrepancy at low energy, it gives rises to very promising results as
documented in the example of elastic scattering observables at 65~MeV
in Fig.~\ref{elast2_65mev}. The NLO predictions are shown as a band
which covers the weak dependence on the choice of the cut-off parameter
$\Lambda$. One sees  clear deviations for $d\sigma/d \Omega$ and the
nucleon analyzing power
$A_y$. Turning to the next order NNLO, which includes the 3N-forces
adjusted as  described above, the agreement for those two observables
is very good.  Also for the tensor analyzing powers $T_{20}$, $T_{21}$,
and $T_{22}$ the agreement  improves. But here unfortunately, the
quality of the only existing data is by far too poor to allow a
definite conclusion. It is a further important  example where precise
and new data are very much needed in order to  put the theory on a
critical test. It will be very interesting to proceed  with this
systematic approach for the study of 3N-force effects and to include
furthermore systematically relativistic corrections showing up  as
additional NN- and 3N-force contributions.

Besides elastic nd-scattering also the nd-break-up process will be a
treasure to probe the nuclear forces  because it has an overwhelmingly
rich structure of break-up configurations leading to a five-fold
differential cross section, and of course, a great variety of spin
observables. While special break-up configurations have  been measured
in the past, sometimes with spectacular  agreement between
theory and experiments, sometimes with nagging  discrepancies,
predictions  have been worked out for regions of phase space where
3N-force effects are  especially large for cross sections as well as
spin observables (up to  factors of 2 to 3).  It appears that
data for the break-up provide the most complete information  on
3N-dynamics and would be very much needed.
 
 \begin{figure}[tbp]
  \centering
  \includegraphics[angle=-90,scale=.4]{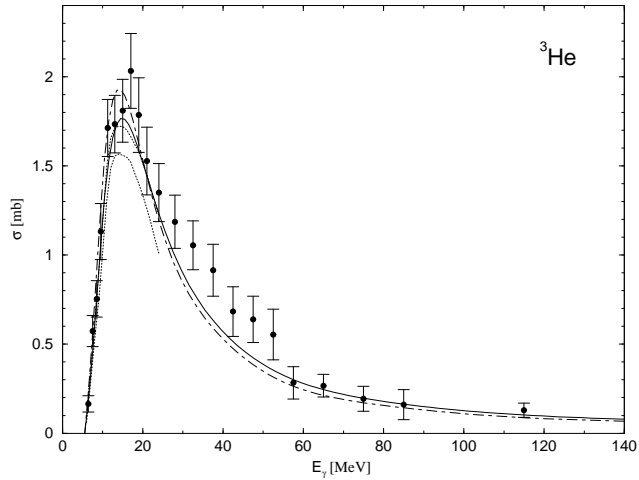}
  \caption{Total photoabsorption cross section of $^3$He in the LIT
approach: AV18 (dash-dotted) and AV18 + Urbana IX (full
curve).}
  \label{gamma_tot_he3_exp}
 \end{figure} 

Likewise the electromagnetic probe represents a valid instrument to
study the 3N-force. First evidence that an electromagnetic reaction is
sensitive to the three-body interaction was shown in the calculation of
the total photoabsorption cross section of $^3$H and $^3$He with
realistic NN- and 3N-forces through the Lorentz integral transform (LIT) 
method. It was shown that 3N-force effects of the order of 10~\%, resulting 
mainly from the change in the triton binding energy, were present at the dipole
resonance peak and in the tail below the pion threshold. The accuracy
of the results was later tested against a FY calculation. It turned out
that the theoretical precision is much higher than the precision of
available  experimental data. As shown in Fig.~\ref{gamma_tot_he3_exp}
even no definite conclusions can be drawn with respect to the interplay
between the NN- and 3N-forces and on the  corresponding two- and
three-body currents as long as new high-precision experimental data is
not available. 

Exclusive reactions (two- and three-body break-up) induced by real and 
virtual photons are very likely more sensitive to 3-body dynamics.
Therefore, precise experimental data on two- and three-body
photodisintegration  cross sections as well as separated (inclusive,
semi-inclusive as well  as completely exclusive) electrodisintegration 
cross sections at lower momenta are needed. The availability of
calculations  which take into account FSI in a complete and accurate
way removes the necessity to restrict the kinematics to the
quasi-elastic  region. Thus it opens up the possibility to study the
transition  from below to that regime. For instance, the spin-dependent
momentum  distribution of polarized proton-deuteron clusters in
polarized $^3$He have  been studied theoretically
via the processes 
$^3 \vec{\mathrm{H}}\mathrm{e} ( e, e'\vec{p}\,)d$ or 
$^3 \vec{\mathrm{H}}\mathrm{e} ( e, e'\vec{d}\,)p$. It turned out that
only at small deuteron momenta ($p_d  \leq 1-2 $ fm$^{-1}$) and for
energies below the pion threshold the process  provides direct insight
into this momentum distribution. For larger momenta,  the longitudinal
and transversal response functions are affected  too strongly by FSI and
MEC's. But the latter effects are equally  interesting, and their
theoretical description should
 be  checked by precise data. Exactly
these type of data and their confrontation with theory will reveal
whether the complex hadron dynamics is theoretically understood.

\subsubsection{The Four-Body Continuum}\label{four_body_cont}
In the quest to go beyond three-nucleon scattering using {\it ab
initio} calculations, four-body dynamics in the continuum is really
a challenging issue. One major goal involves an answer to the following
question: can we understand the simplest four-nucleon
scattering problem using present NN- or NN+3N-force models as we do
for most $nd$ and $pd$ scattering observables?  The A = 4
system is the lightest nuclear system displaying  clear cut resonances
in physical observables. One of the best examples is the total
neutron-triton cross-section which shows a broad resonance peak
centered at $E_n = 4$ MeV. Despite its nicely looking form, a careful
R-matrix analysis reveals overlapping broad resonances.
None of the related scattering phase shifts passes ever through
$90^\circ$.  This is a common feature of all (A=4)-nuclei
$^4$H, $^4$He, and $^4$Li. 
 
The study of $n + ^3$H elastic scattering (or $p + ^3$He) below three-body
break-up threshold is the simplest 4N-scattering problem because its
total isospin is $I = 1$, and up to $E_n = 8.34$ MeV the problem reduces
to the elastic channel alone. In the past five years a number of
calculations emerged using different methods or force 
models. Although some small disagreement with data shows up
at threshold energies, no fully converged exact calculation
using realistic NN-interactions is able to describe the total $n + ^3$H
cross section at the resonance peak ($E_n = 4$ MeV). In the near future
more calculations applying different NN-forces are expected and studies are underway to understand the role of 
the 3N-force and the sensitivity to changes in NN-force models. In
addition to accurate benchmark  calculations to clarify
possible discrepancies between the  results of different groups, one
needs precision neutron data for scalar and vector observables. The
absence of Coulomb force in $n + ^3$H scattering opens a realm of
theoretical possibilities to the solution of the 4N-problem which
leads to the possibility of using a wide variety of (2N + 3N)-force
models. 

\begin{figure}[tbp]
\centering
\includegraphics[scale=.4]{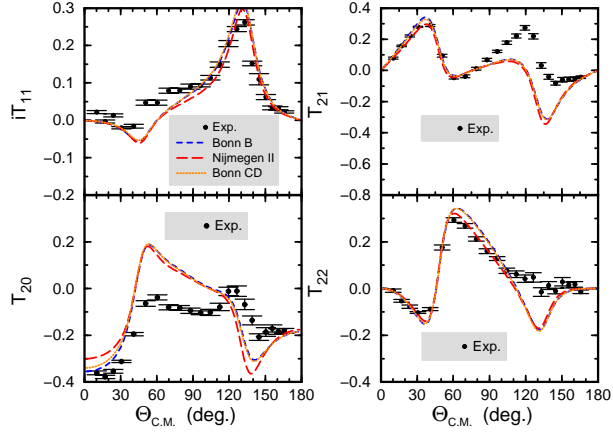}
\caption{Tensor analyzing powers for $dd \to pt$ at $E_d = 6.1$ MeV for
different NN-potentials.}
\label{T20-iT11}
\end{figure}

Using the AGS equations and a rank one separable representation of
realistic NN-interactions, a complete study of all four-nucleon
reactions has been made at energies up to the four-body
break-up threshold. These calculations uncover large
disagreements in $dd \to dd$ and $dd \to p^3$H tensor analyzing
powers as shown in Fig. \ref{T20-iT11}, as well as an $A_y$ deficiency
in $n^3\mbox{He} \to n^3\mbox{He}$ and $n^3\mbox{He} \to p^3$H.
Nevertheless the total cross section data for $n + ^3$H elastic
scattering in the resonance peak is well accounted for by these
calculations. Confirmation of these findings requires the use of the
full NN-potential.
 
\begin{figure}[tbp]
\centering
\includegraphics[scale=.4]{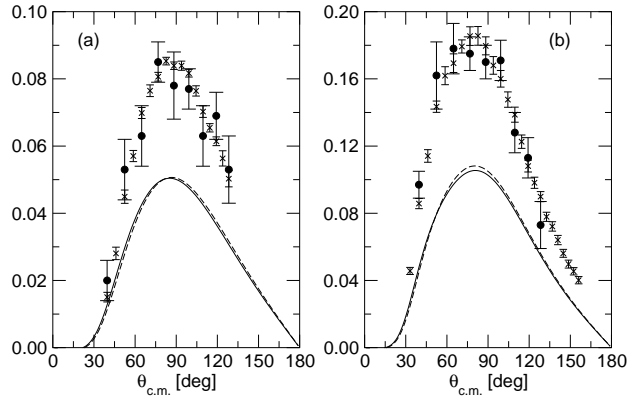}
\caption{Vector analyzing power $A_y$ versus $\theta_{CM}$ for $\vec
p + ^3$He reaction at $E_{CM}~=~1.2$~MeV and 1.69 MeV respectively.
The solid lines (dashed lines) correspond to calculations with AV18
alone while the dashed lines to AV18+ Urbana IX.}
\label{viviani-ay}
\end{figure} 

\begin{figure}[tbp]
\centering
\includegraphics[scale=.4]{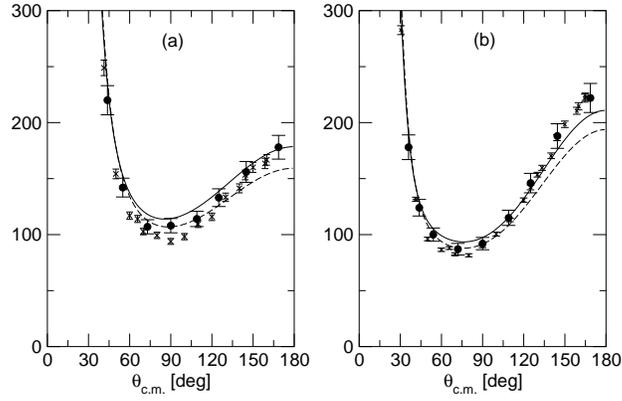}
\caption{Differential cross section for the same reaction as in 
Fig.~\ref{viviani-ay} with the same notation.}
\label{viviani-xsu}
\end{figure} 

\begin{figure}[tbp] 
\centering
\includegraphics[scale=.7]{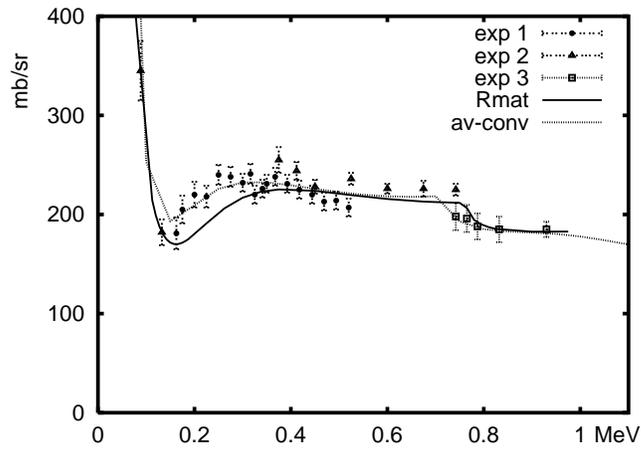}
\caption{Comparison of various data for $p-^3$H scattering at 120 degrees
with  the R-matrix analysis (solid) and the AV18 
calculation (dotted).}
\label{h2} 
\end{figure}

Like in $p + d$ elastic scattering, fully converged variational
calculations also find an
$A_y$ discrepancy in $p + ^3$He scattering at low energy
(Fig.~\ref{viviani-ay}), but in addition  a fairly large disagreement
with the existing data for the differential cross section
(Fig.~\ref{viviani-xsu}). One also finds that
$p+^3$He scattering lengths extracted from effective range phase shift
analysis depend on the data sets one includes. In particular if cross
section and analysing power measurements at proton energies below 1 MeV
are added to the higher energy data ($E_p < 12$MeV), one gets two sets
of solutions. This is a clear indication that one may need high
precision data at low energy to clarify the situation. 

\begin{figure}[tbp]
  \centering  
\includegraphics[scale=.7]{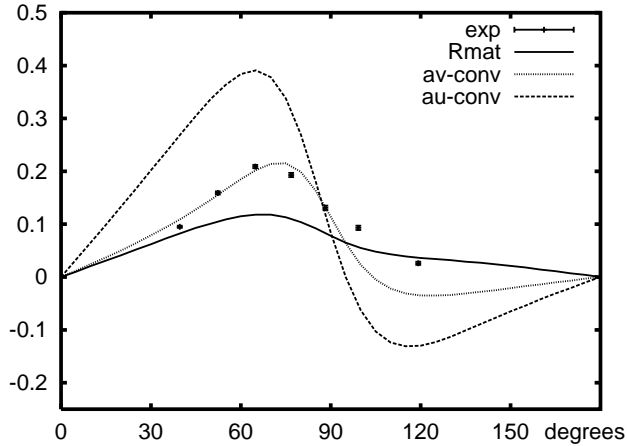}
  \caption{Predictions of proton analyzing power in  $^3{\rm
H({\vec{p}},n){^3He}}$  at 3 MeV $E_{c.m.}$ from R-matrix analysis and
calculations using AV18 alone (av-conv) and AV18 together with 
Urbana IX  (au-conv). The data shown at 2.9~MeV is very energy dependent.}
  \label{h3}
\end{figure}  

In the $p + ^3$H sector one finds very old data such as shown in
Fig.~\ref{h2}. This situation makes it impossible to determine the
position of the related $0^+$ resonance which enters into the
determination of the nuclear compressibility coefficient. Another
example is the charge exchange reaction $^3$H(p,n)$^3$He or its inverse
reaction. Here again the vector polarizations are very sensitive to the
3N-force as seen in Fig. \ref{h3}. For the corresponding elastic
scattering cross sections the sensitivity is quite reduced. The reason
for this  feature can be traced back to the isospin structure of
various partial waves. The $0^+$ and $0^-$ transition matrix elements
reach almost the unitary limit of one, forcing  the elastic one close
to zero. Although this may lead to a reduced  sensitivity in the
elastic channels, one challenging issue involves the correct
determination of the position and width of the $0^-$ resonance in $I =
0$. 

Alternative approaches to the quite challenging problem of four-body 
dynamics in the continuum are provided by integral transform methods.  
The LIT method is particularly suited for inclusive reactions, which 
are very complicated to compute in the traditional way due to the
increasing  number of contributing channels with increasing particle
number. First applications have been considered for e.m.\ inclusive
reactions,  although promising results for exclusive as well as purely
hadronic  reactions have already been obtained in a simple two-nucleon
test case. Particularly interesting are the results of a LIT calculation
for the longitudinal $^4$He $(e,e')$-response function at various
momentum transfers between 300 and 600~MeV/c using a semi-realistic NN
potential. The theoretical results could show that the effect of FSI is rather 
large, even in the quasi-elastic peak region. The FSI effect decreases with 
increasing momentum transfer, but nonetheless it became clear that 
quasi-elastic approaches are justified only at momenta  well beyond 500~MeV/c 
and only at peak kinematics. It was also found that kinematics at low energies 
and relatively high momenta, which were suggested in order to study bound 
state correlation effects are not at all exempt from FSI. The obtained results
are in good agreement with experimental data (size of experimental error
about 10$\div$20 \%). It remains to be seen if the good agreement, in 
particular at  high energy and momentum, survives a more realistic calculation 
which includes also lowest order relativistic contributions and more precise 
data. We would like to point out that with higher quality $^4$He$(e,e')$-data
one  could study more specific questions, e.g., the importance of
3N-forces. Preliminary studies on 3-body systems indicate
that  3N-forces reduce the quasi-elastic peak height by about 10~\% and
increase the high-energy tail by about the same amount. Such effects
could be even larger in the four-body system, but one will certainly
need more precise data than the present ones in order to obtain a clear
picture. 

Additional important insights into the hadronic structure of the 4N-system 
could be found studying the inclusive transverse $(e,e')$-response, e.g., 
about the role of MEC. Clear evidence of 
their importance as well as surprising differences between the case of 
$^4$He and those of 3- and 6-body systems have been seen using another 
integral transform approach analysing Euclidean responses. The results
show that MEC contributions  seem to be particularly amplified in $^4$He 
and increase considerably with decreasing momentum transfer. Thus one can 
certainly state that a separation of longitudinal and transverse responses
in inclusive electron scattering experiments on $^4$He at lower momentum 
transfers would be very valuable. Important aspects of the hadronic 
force and questions related to the consistency of the MEC with such a force
could be investigated. 

 \begin{figure}[tbp]
  \centering
  \includegraphics[angle=270,width=.65\textwidth]{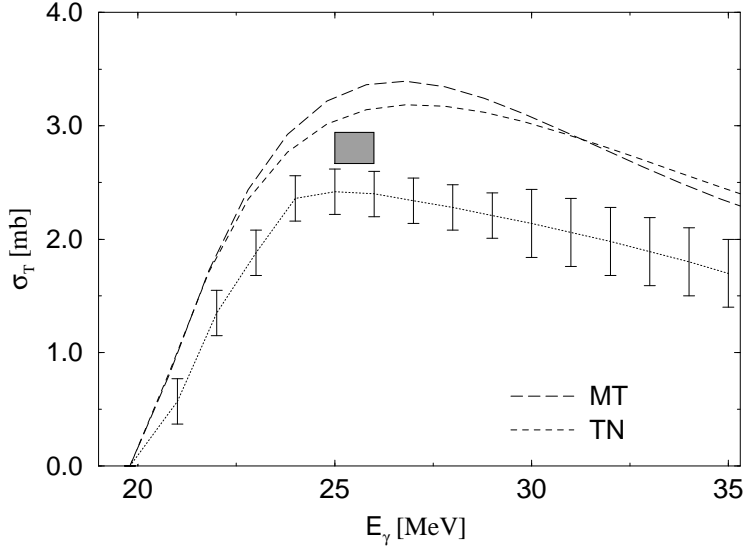}
  \caption{Total photoabsorption cross section of $^4$He 
obtained with the LIT method with
TN and MT potentials; experimental results:
sum of $(\gamma,n)$ and $(\gamma,p)^3$H cross
sections (dotted curve with indication of error range) and indirect
determination via Compton scattering (shaded area).}
  \label{gamma_tot_he4}
 \end{figure}
 
Besides the inclusive $^4$He$(e,e')$-reaction, the two-body break up 
channels should be studied, also including polarization effects
(electron and/or polarization of outgoing particles). Such studies could
reveal other important details of the hadronic structure than seen in 
inclusive $(e,e')$ or purely hadronic four-body reactions. First 
investigations have already been carried out. Complete agreement between
experiment and theory has not yet been obtained.
For future studies the region of low momentum transfer is especially 
interesting.
In particular, one could investigate the transition from virtual 
to real photon physics. We would like to mention that with respect to
the latter, striking  differences between theory and available old
experimental data have been found as seen in Fig.~\ref{gamma_tot_he4}
and need further investigation both theoretically and experimentally.
An experimental program has already started at Max-Lab in Lund. Other basic
properties like electric polarizabilities and magnetic susceptibilities
are largely unknown. They are accessible both via Compton scattering or
via integrated photoabsorption cross sections to the extent that one is
able to separate electric and magnetic processes. Their study would be
interesting not only as a test of the underlying forces, but also to get
an intuitive representation of their structure.




\section{Microscopic Description of Many-Body Phenomena}
\label{many_body_systems}

As we have seen above, very light nuclei ($A=2\div 4$) 
are privileged systems for the study of fundamental issues like properties
and origin of the strong force. However, the study of heavier nuclei 
has its own merit with respect to new many-body phenomena like, e.g.,
clusterization and collective motion. Furthermore, one can expect that 
in heavier systems some of the interaction phenomena found in light nuclei
may be modified or amplified in view of the higher average nucleon density 
(``medium effects''). 

One of the principal advantages of studying very light systems 
resides in the fact that calculations are accurate enough in order to make the 
comparison with precise data conclusive with respect to the issues 
considered above. Therefore, in order to reach the same conclusive 
strength for heavier nuclei, it is important to produce also for systems 
with $A>4$ on the one hand accurate theoretical results and on the other 
hand accurate experimental data.

\subsection{Properties of Ground and Low-Lying Excited States}

Going beyond four-body nuclei, one finds quite significant effects from
the 3N-force. Allowing for additional spin-isospin 
dependences in the 3N-force, leading to the model IL4,
in addition to $^3$H and $^4$He the low-lying spectra for up to A=8 nucleons
could be rather well described as shown in Table~\ref{table4}. 
These results clearly show that, in relation to the 
most modern NN-forces, 3N-forces are unavoidable in order to describe 
binding energies and low-lying excited states of light nuclei. 
Since the number of nucleon triplets overtakes more and more the number
of nucleon pairs with increasing A, it is clear that 3N-forces 
have to be included in any realistic description of complex nuclei, too.
\begin{table}[tb]
\caption{
Experimental and GFMC energies (in MeV) of particle-stable 
or narrow-width nuclear states. 
Monte Carlo statistical errors 
in the last digits are shown in parentheses.
}
\begin{center}
\begin{tabular}{lrrr|lrrr}
\hline
                      & AV18       &  IL4       &  Exp  &
                      & AV18       &  IL4       &  Exp  \\
\hline
$^3$H  ($\frac12^+$)  &  -7.61(1)  &  -8.44(1)  & -8.48 &
$^7$Li ($\frac12^-$)  &  -31.1(2)  &  -39.0(2)  &  -38.77 \\
$^3$He ($\frac12^+$)  &  -6.87(1)  &  -7.69(1)  &  -7.72 &
$^7$Li ($\frac72^-$)  &  -26.4(1)  &  -34.5(2)  &  -34.61 \\
$^4$He ($0^+$)        &  -24.07(4) &  -28.35(2) &  -28.30 &
$^8$He ($0^+$)        &  -21.6(2)  &  -31.9(4)  &  -31.41 \\
$^6$He ($0^+$)        &  -23.9(1)  &  -29.3(1)  &  -29.27 &
$^8$Li ($2^+$)        &  -31.8(3)  &  -42.0(3)  &  -41.28 \\
$^6$He ($2^+$)        &  -21.8(1)  &  -27.4(1)  &  -27.47 &
$^8$Li ($1^+$)        &  -31.6(2)  &  -40.9(3)  &  -40.30 \\
$^6$Li ($1^+$)        &  -26.9(1)  &  -32.0(1)  &  -31.99 &
$^8$Li ($3^+$)        &  -28.9(2)  &  -39.3(3)  &  -39.02 \\
$^6$Li ($3^+$)        &  -23.5(1)  &  -29.8(2)  &  -29.80 &
$^8$Li ($4^+$)        &  -25.5(2)  &  -35.2(3)  &  -34.75 \\
$^7$He ($\frac32^-$)  &  -21.2(2)  &  -29.3(3)  &  -28.82 &
$^8$Be ($0^+$)        &  -45.6(3)  &  -56.5(3)  &  -56.50 \\
$^7$Li ($\frac32^-$)  &  -31.6(1)  &  -39.5(2)  &  -39.24 &
$^8$Be ($1^+$)        &  -30.9(3)  &  -38.8(3)  &  -38.35 \\
\hline
\end{tabular}
\label{table4}
\end{center}
\end{table}

\begin{figure}[htb]
\centering
\includegraphics[angle=-90,width=.65\textwidth]{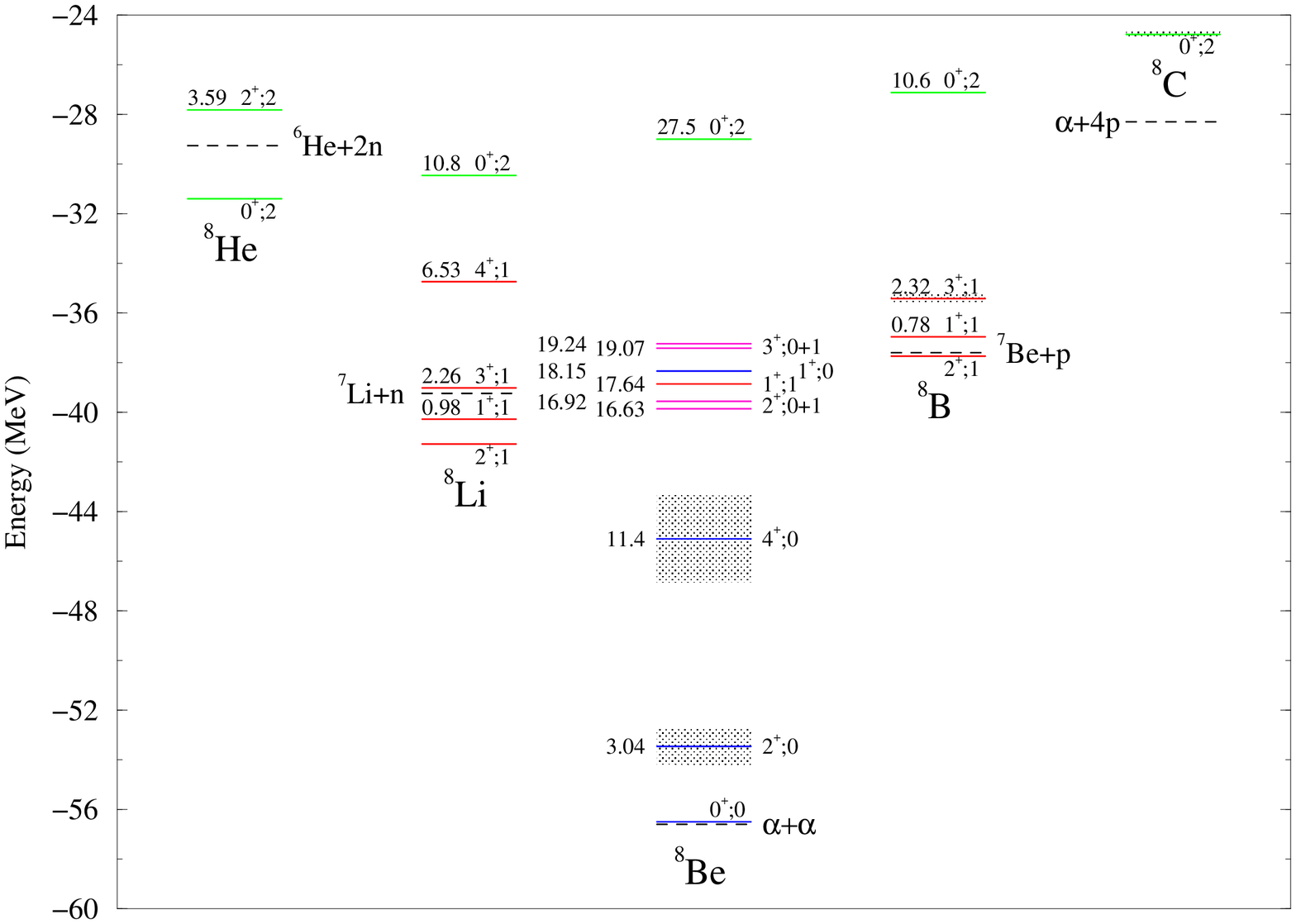}
\includegraphics[width=.65\textwidth]{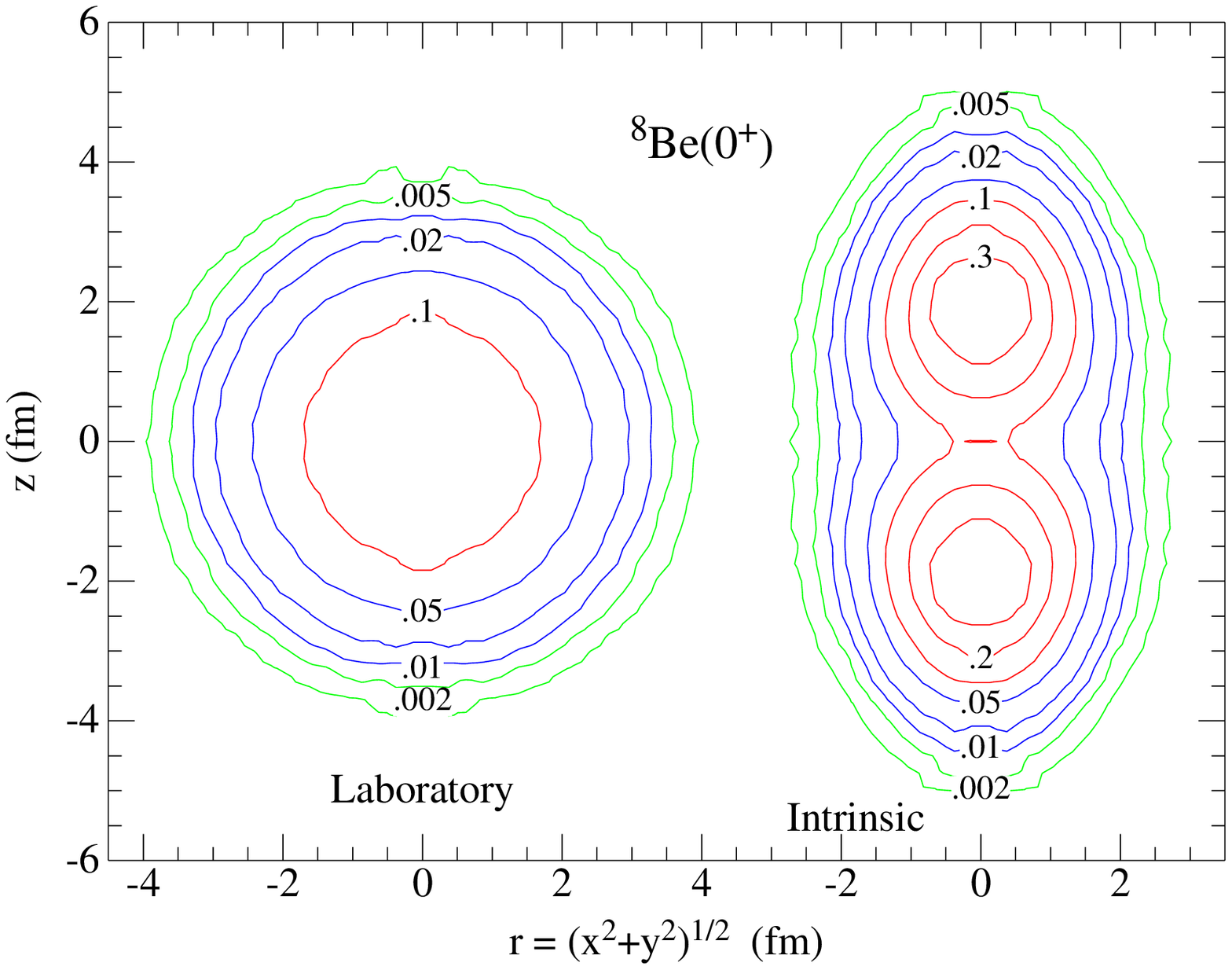}
\caption{
Upper panel: calculated spectrum of $^8$Be ;
lower panel: calculated density contours of the $^8$Be ground state 
in the lab frame(left) and 
the intrinsic frame (right), 
labeled with densities in fm$^{-3}$.
}
\label{be8spectrum}
\end{figure}

Already for A~$=4,\, 6,\, 8,\dots$ 
one may observe phenomena which are precursors of 
the above mentioned many-body phenomena. To illustrate this point of view, 
we show in the upper panel of 
Fig.~\ref{be8spectrum} the spectrum of $^8$Be as obtained in a 
microscopic ``Variational Monte Carlo'' - ``Greens Function Monte Carlo'' 
calculation (VMC-GFMC) which 
is based on one-body orbitals with four nucleons in an $\alpha$-core 
coupled to (A-4) one-body ($\ell=1$) wave functions. 

The low-lying states of $^8$Be resemble a rotational spectrum which 
in fact can be described quite nicely in a cluster-model where
two $\alpha$-clusters rotate around their common center of mass.
However, it is also possible to recover this picture from the VMC wave 
functions by a modified Monte Carlo density calculation.
This can be seen in the lower panel of 
Fig.~\ref{be8spectrum} which shows contours of constant 
density plotted in cylindrical coordinates. The left side of this panel
shows the lab frame density. For the $J$~=~0 ground state it has to be 
spherically symmetric. On the other hand, the intrinsic density shown 
on the right side exhibits clearly two peaks, with the neck
between them having only one-third of the peak density. It is obvious
that this feature can be interpreted as the clusterization into two 
$\alpha$'s of $^8$Be.     

This interpretation is corroborated by looking at the densities 
of the $J=2^+$ and $4^+$ states. Although 
the laboratory densities for these states (in $M=J$ states) are
quite different, the intrinsic densities are, within statistical errors,
the same as the $J=0$ intrinsic density, which is typical
for a rotational band. In fact, if the $0^+$, $2^+$, and $4^+$ states are 
generated by rotations of a common intrinsically deformed structure, 
then their electromagnetic moments
and transition strengths should all be related to the intrinsic
moments which can be computed by integrating over the
projected body-fixed densities.  This has been explored,
and a consistent picture is obtained.

 \begin{figure}[tbp]
  \centering
   \includegraphics[width=0.7\textwidth]{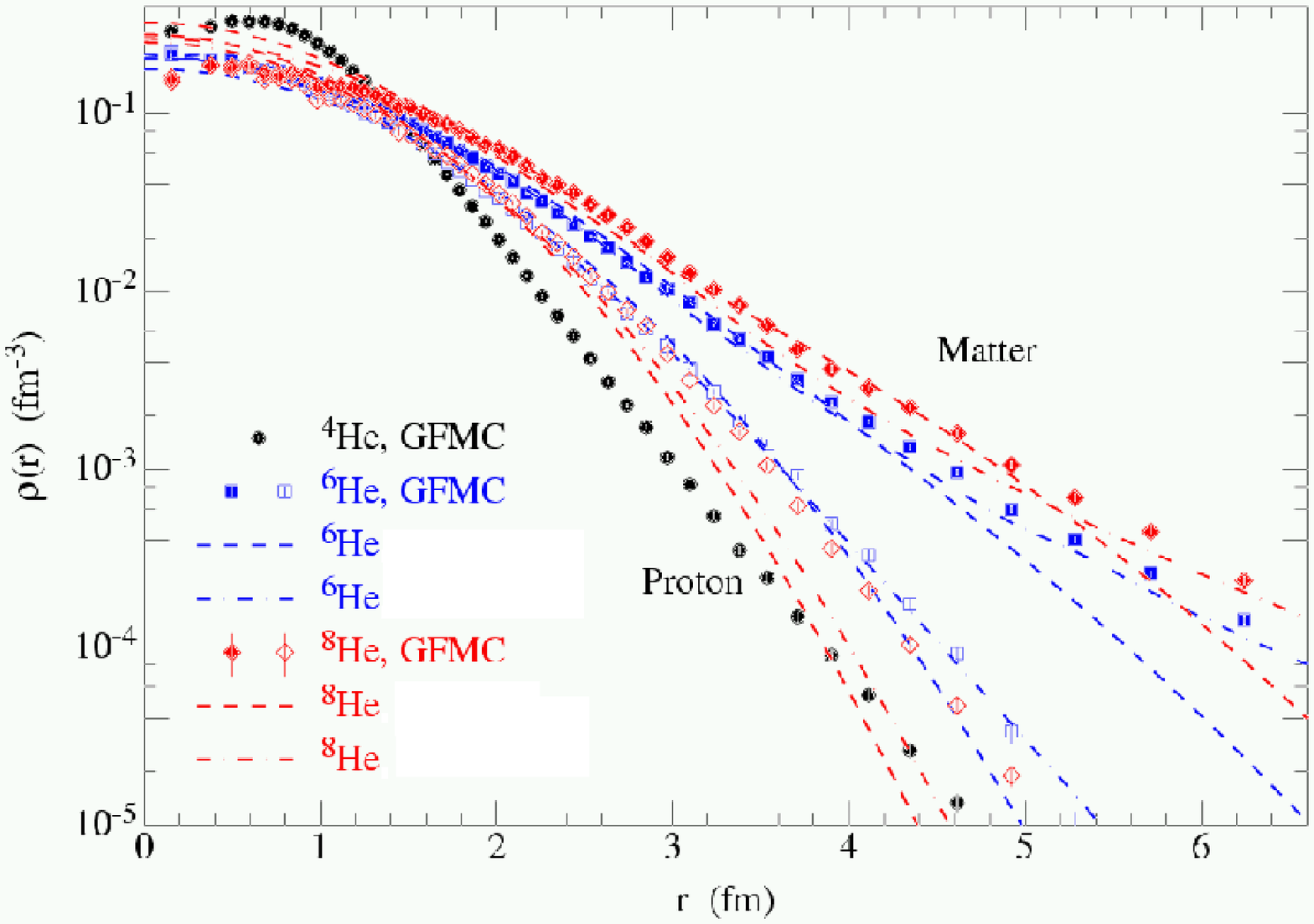}
  \caption{Proton and matter densities in $^4$He, 
  $^6$He and  $^8$He. Analysis of $^{6,8}$He proton scattering are shown by the dashed and dot dashed curves.}
 \label{he-rho1}
 \end{figure}

The {\it ab initio} quantum Monte Carlo calculations are also used to study 
the structure of some of the lightest halo nuclei like $^{6,8}$He. 
Fig.~\ref{he-rho1} shows predicted densities compared to those deduced 
from experiment. It can be seen
that, as more neutrons are added, the tails of the matter 
distributions broaden considerably because of the relatively weak binding
of the $p$-shell neutrons. In addition, the central neutron and proton 
densities decrease rather dramatically. This effect does not
necessarily require any changes to the $\alpha$-core but can be understood 
at least partially from the fact that the $\alpha$ no longer
sits at the center of mass of the entire system. The motion relative 
to the center of mass spreads out the mass distribution relative
to that of $^4$He. 
Proton-proton distribution functions have also been calculated and one can 
compare the ones of $^4$He to those of $^6$He and $^8$He. 


 

If the $\alpha$-core of $^{6,8}$He is not
distorted by the surrounding neutrons, all three $pp$ distributions 
should be the same. One sees, however, that the $pp$ distribution
spreads out slightly with neutron number in the helium isotopes, 
with an increase of the pair r.m.s. radius of 4\% in going from $^4$He to 
$^8$He.
Although this could be interpreted as a swelling of the $\alpha$-core, it 
might also be caused by charge exchange correlations. 
This needs to be further investigated

Due to the complexity of complete many-body calculations for larger and
larger nuclei one has often resorted to the much simpler
descriptions in terms of clusters that grasp the essential physics 
of a given problem. The methods normally applied to few-nucleon systems 
are then applied to few-cluster systems. It is highly interesting to
test cluster-model results to A-body calculations in order to be aware of 
the range of applicability and the limits of these descriptions of many-body 
systems. These comparisons are now possible. 
The VMC wave functions for the AV18-Urbana IX model has been used to
calculate a variety of cluster-cluster overlap wave functions, such as
$\langle d p | t \rangle$, $\langle d d | \alpha \rangle$, and $\langle
\alpha d | ^6{\rm Li} \rangle$.

Recently also the overlaps 
$\langle^6{\rm He}(J^{\pi})+p(\ell_j)|^7{\rm Li}\rangle$ for 
all possible $p$-shell states in $^6$He were studied.
The spectroscopic factors, obtained from these overlaps, are 0.41 to the 
ground state of $^6$He and 0.19 to the 2$^+$ first excited state.
These factors are significantly smaller than the predictions of the 
Cohen-Kurath (CK) 
shell model, which gives values of 0.59 and 0.40, 
respectively.
The CK shell model requires that the possible $^6$He$+p$ states sum to unity
within the $p$-shell, whereas in the VMC calculation, the correlations in 
the wave function push significant strength to higher momenta that cannot 
be represented as a $^6$He state plus $P$-wave proton.

Proceeding with this kind of investigations to larger A will allow one
to show where and whether the transition to mean field structures will 
show up. It will be very interesting to see whether spectroscopic factors 
close to unity will at some point appear as a result of microscopic
A-body calculations and single-particle features which may appear
in  one-body knock-out experiments will then have an interpretation
in terms of a more fundamental theory.

\begin{figure}[tbp]
  \centering
   \includegraphics[width=0.5\textwidth,angle=270]{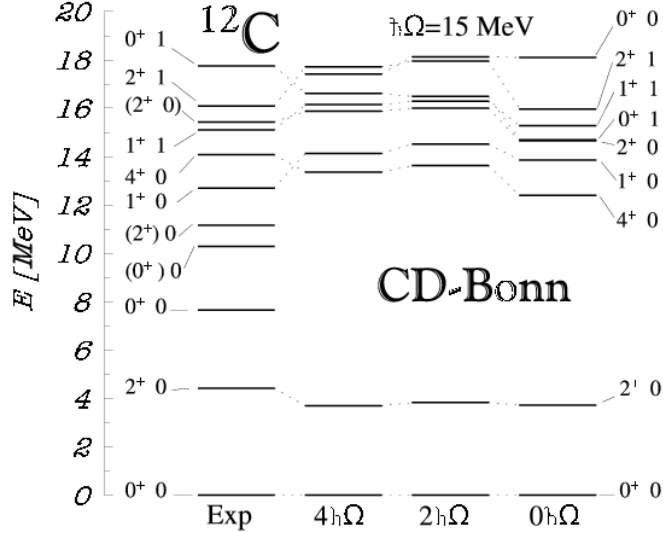}
  \caption{Experimental and theoretical positive parity excitation spectra 
   of $^{12}$C}
 \label{navratil}
 \end{figure}

First calculations in the region of larger A up to
A=16 have already been carried out with VMC predictions of 
spectroscopic factors 
and NCSM calculations of the low-lying spectra and some
electromagnetic properties. 
In particular the NCSM method, which has proved to be
very accurate in calculating ground state properties of the traditional light 
systems (see e.g. Table 1), has also been pushed to study the low-lying 
spectrum 
of $^{12}$C (see Fig.~\ref{navratil}) and recently even the Gamow-Teller excitations of A=14 
nuclei.

Though at present the calculations  are not fully convergent they represent
the first examples of {\it ab initio} calculations with realistic interactions
in a mass region which has been studied extensively for many years 
with  simplified and approximate methods. Thus one may hope that in the 
near future the microscopic interpretation 
of many-body phenomena will be available.     

\subsection{Continuum States and Collective Motion}

Unambiguous interpretations of experiments in terms of nuclear structure 
properties require in most cases a reliable treatment of the continuum. 
Inelastic processes in fact constitute a much richer source of information.
Indeed, if one wants to get at single-particle properties like, e.g., 
shell momentum distributions via one-body knock-out experiments, or to 
study the microscopic origin of collective motion, one needs to have 
the continuum under control.

 \begin{figure}[htb]
  \centering
  \includegraphics[angle=-90,width=1.0\textwidth]{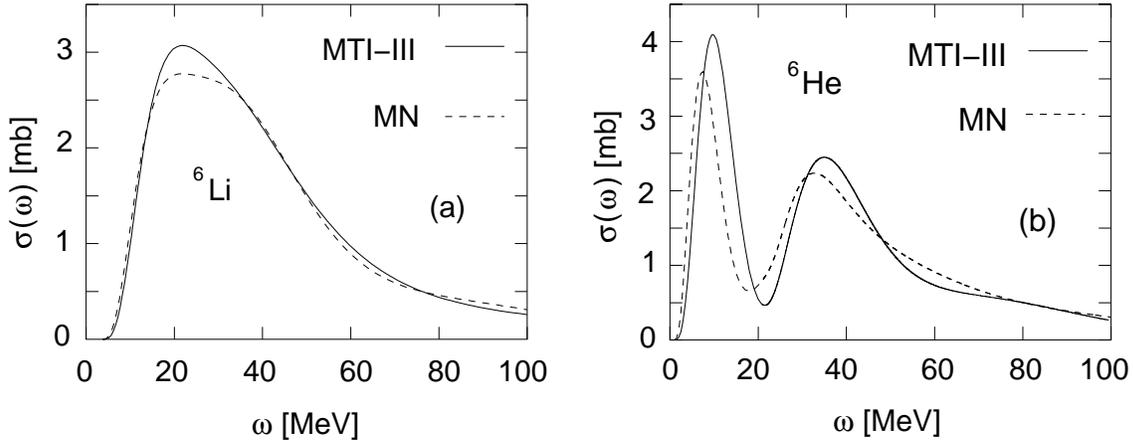}
  \caption{
Total photoabsorption cross section of the $^6$Li and $^6$He
      in the LIT approach  with MT  and Minnesota potentials.
  } 
  \label{6body}
 \end{figure}

 \begin{figure}[htb]
  \centering
\centerline{\psfig{figure=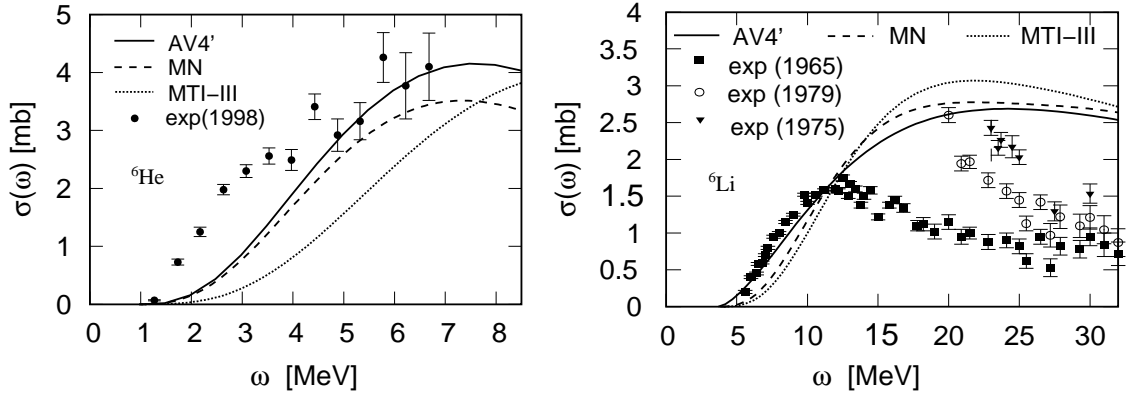,width=1.0\textwidth}}
  \caption{
As Fig.~\ref{6body} with additional experimental data.
  } 
  \label{6body_exp}
 \end{figure}
In this respect considerable progress has been achieved in recent 
calculations of reactions on 6-body systems using integral transform 
methods. In fact, a combination of the LIT and EIHH 
methods allows one to calculate inclusive cross sections on a full A-body 
microscopic basis. An example is shown in Fig.~\ref{6body} displaying the 
results of the photoabsorption cross section $\sigma(\omega)$ of
$^{6}$Li and $^{6}$He . 
The very interesting result of this calculation is the appearence of a single 
giant dipole resonance peak for $^{6}$Li in contrast to two well separated 
peaks for $^{6}$He. One can associate the low-energy peak of $^{6}$He at 
$\omega \simeq 8$ MeV to the break-up of the
neutron halo, whereas the second one, at about $\omega$=35 MeV, should 
correspond to 
the break-up of the $\alpha$-core. Interpreting this feature as a collective 
motion, one can distinguish a soft collective mode 
due to oscillations of the core against the halo part of the nucleus 
from a classical Goldhaber-Teller mode of the proton sphere against 
the neutron sphere. On the other hand, the $^{6}$Li cross section 
does not show such a substructure. This is probably due to
the fact that the break-up into two three-body nuclei, $^{3}$He~+~$^{3}$H, 
fills the gap between the two modes. Note that 
in case of $^{6}$He a corresponding break-up into two identical nuclei,
$^{3}$H~+~$^{3}$H, is not induced by the dipole operator. 
The comparison with experimental data in Fig.~\ref{6body_exp} shows that 
the theoretical cross sections with semirealistic central $S$-wave
interactions (MT, Minnesota) miss some strength at very low energies. 
Inclusion of $P$-wave interaction (AV4') leads to much improved results
for $^{6}$Li.  There is only a small effect in $^{6}$He, contrary to
what is obtained in a cluster-model
with an inert $\alpha$-core and two neutrons interacting via a 
$P$-wave potential.   
The situation in the
region of the giant dipole peak is much less clear, either because of lack of 
experimental data ($^6$He) or because existing data do not lead to a 
unique picture ($^6$Li).

This example shows how important and urgent it is to intensify on the 
one hand the theoretical investigations of the microscopic dynamical 
structure of such many-body systems, and on the other hand to start 
or at least to renew an experimental program for obtaining accurate 
data for such targets both for the inclusive process and for the 
various exclusive channels as well. To give an example, it 
would be particularly interesting to see whether the $^3$He-$^3$H 
channel has a peak in between the soft and hard modes. Of course, 
a rich field would be opened by detailed studies of the deformed structure 
of such nuclei in terms of collective motion, like in the case of $^8$Be.




\section{Few-Body Physics Related to Other Fields}\label{few_body_slave}

\subsection{Two- and Three-Body Systems as Effective Neutron Targets}
\label{neutron_target}

Another very important aspect of few-body nuclear physics is the use of
lightest nuclei like deuteron or $^3$He as effective neutron targets 
in view of the fact that free neutron targets do not exist. The method
rests on the assumption that (i) the binding is weak, and (ii) 
FSI effects and other hadronic influences due to the 
presence of spectator nucleons can either be neglected or reliably 
corrected for. This means that only a full control of the nuclear few 
body dynamics allows one to obtain clear cut information on the 
intrinsic properties of the neutron.

A typical example is the quest for the electromagnetic 
form factors of the neutron, in particular the small electric one $G_{En}$
in quasifree electron scattering on deuterium or 
$^3$He, exploiting polarization 
observables. For deuteron electrodisintegration a systematic 
study of polarization observables has been performed 
with respect to kinematical regions where 
effects of two-body dynamics from FSI and MEC were 
minimal and on the other hand the dependence 
on $G_{En}$ was maximal with the intention to guide experimental efforts.
It turns out that especially for low momentum transfers the medium 
effects are sizeable, meaning that the determination of $G_{En}$ 
relies heavily on the theory for the evaluation of such medium effects
in a reliable manner. In fact the determination of $G_{En}$ is done in 
such a way that one calculates the corresponding observables for varying
values of $G_{En}$ and comparing the result with the experimental data.
An interesting check of this method is provided by performing the 
corresponding experiment on a bound proton. Similar theoretical studies 
for $^3$He have been done and
promising results have also been achieved using a polarized $^3$He target
for the determination of $G_{En}$. 

 \begin{figure}[tbp]
  \centering
  \includegraphics[scale=.4]{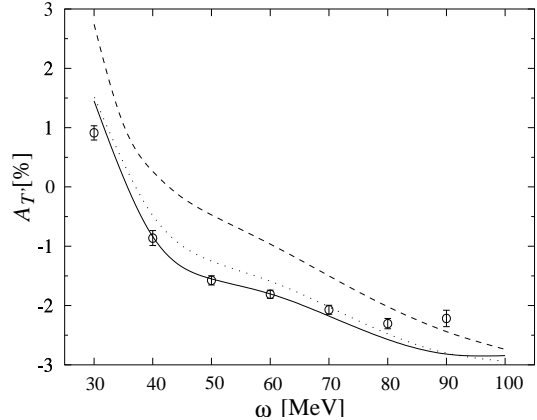}
  \caption{The transverse asymmetry $A_{T'}$ of quasi-elastic electron 
scattering on $^3$He at $Q^2=0.1$~(GeV/c)$^2$. Dashed:
PWIA; dotted: FSI included; solid: FSI+MEC.
  }
  \label{fig7}
 \end{figure}
Another example, shown in Fig.~\ref{fig7}, is the asymmetry $A_{T'}$ 
for quasi-elastic scattering of polarized electrons on polarized 
$^3$He, which is known to be sensitive to the magnetic form 
factor of the neutron $G_{Mn}$. This very nice agreement with the data 
allowed one to extract $G_{Mn}$ in perfect agreement with the value 
extracted on the deuteron using the ratio of the asymmetries related to the 
processes $d(e,e'n)$ and $d(e,e'p)$. It should be emphasized 
again that this was only possible because FSI and MEC's were properly taken 
into account. 
 
\subsection{Few-Body Reactions of Astrophysical Relevance}
The control of the continuum allows one to study also electroweak processes
of astrophysical relevance. This is of particular importance for those
reactions whose cross sections are either impossible or very difficult 
to measure in the laboratory. As an example, we mention the $pp$ 
weak capture $p+p\rightarrow d+e^+ +\nu_e$, the most fundamental process 
for energy production in main-sequence stars, and the hep process 
$p+^3\!\mathrm{He}\rightarrow ^4\!\!\mathrm{He}+e^+ +\nu_e$. The knowledge of 
the reaction rate, therefore, is based on the development of realistic 
models for the nuclear dynamics and the electroweak transition operators.
In fact, a recent calculation in the framework of EFT has improved 
considerably the precision for evaluating the threshold 
$S$-factors of these reactions. 
  
Another example is the ${}^2$H$(p,\gamma){}^3$He capture
reaction. This process is extremely sensitive to the two-body
electromagnetic transition operators and to the small components of
the nuclear wave functions. This is because the
one-body transition operators cannot connect the main $S$-state
components of the deuteron and $^3$He wave functions. 
Accurate data for this reaction exist and, therefore, interesting tests 
can be performed of both the nuclear Hamiltonian from which the
nuclear wave functions are obtained, and the model used to
describe the nuclear currents. Since the nuclear e.m.\ current is 
related to the Hamiltonian through current conservation,
it is clear that the two topics are inter-related. 

 \begin{figure}[h]
  \centering
  \includegraphics[scale=.5]{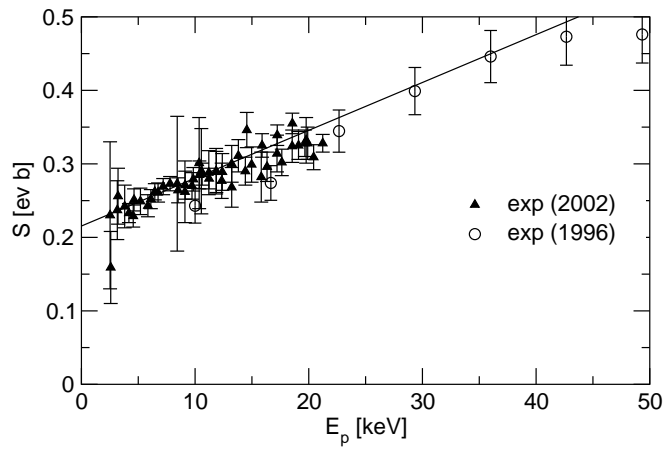}
  \caption{
Astrophysical $S$-factor of the $^2$H$(p,\gamma)^3$He capture reaction 
at low energies. The curve shows the theoretical calculation using the 
AV18+Urbana IX interaction model. Data from TUNL 
and LUNA.
  } 
  \label{LUNA}
 \end{figure}
For illustration we show in Fig.~\ref{LUNA} the calculated and
measured astrophysical $S(E)$-factor at low energy. 
The calculation has been performed using the MEC model derived by the 
Urbana-Argonne collaboration. As one readily notes, a good agreement 
with the LUNA data is found. Also at higher energies, a good agreement 
between theory and experiment is achieved in general, 
but with some notable exception, represented by the tensor analyzing 
powers. These observables are in fact rather sensitive to the 
details of the two-body e.m.\ current and therefore
represent an optimal testing ground for the e.m.\ current model. 

\section{Conclusions}\label{conlusions}

As a summary we may conclude that the achievements in the field of 
few-nucleon physics reached up to now are very promising in the sense 
that one can hope to finally understand the low-energy observables 
in light nuclei quantitatively in terms of the basic hadronic two- 
and three-body forces combined with accurate theoretical tools for 
solving the quantum mechanical few-body problem. Besides properties 
of ground and low-lying excited states, hadronic and electromagnetic 
reactions will provide a great variety of experimental observables 
against which the theory can be checked. Many of the problems
debated in the seventies and the eighties
and left unsolved for lack of theoretical guidance, could be addressed
again nowadays with a more fundamental point of view, which focuses on
subnuclear non-perturbative physics. In view of the in many cases 
very sparse experimental information it is of utmost importance that 
experimental programs in low-energy nuclear physics should be supported
on a broader scale than has been done in the recent past. 

In addition, pushing the above mentioned methods to solve nuclear 
systems with increasing number of particles will give us a much deeper
insight into nuclear many-body phenomena, which 
traditionally are described in shell or cluster-models or in an even 
more phenomenological collective model. In this way
such models would be put on a more solid theoretical ground by establishing a 
quantitative relation to the basic nuclear forces. 
Furthermore, accurate information gathered in 
few-nucleon physics, like spin structures of light nuclei, will be very
useful for the analysis of high-energy experiments.





\addcontentsline{toc}{section}{References}

\begin{thebibliography}{99}


\bibitem{1}
"Few-Body Problems in Physics '95"
Proceedings of the XVth European Conference, Peniscola (Castillon), Spain, 
June 5-9, 1995,
Edited by R. Guardiola, Few-Body Systems, Suppl. 8, 1996. 

\bibitem{2}
"Few-Body Problems in Physics"
Proceedings of the XVth International Conference, Groningen, The Netherlands, 
July 22-26, 1997, 
Edited by J.C.S. Bacelar, A.E.L. Dieperink, L.P. Kok and R.A. Malfliet, 
North Holland Elsevier, 1998.

\bibitem{3}
"Few-Body Problems in Physics '98"
Proceedings of the XVI European Conference, Autrans, France, June 1-6, 1998, 
Edited by B. Desplanques, K. Protasov, B. Silvester-Brac, J. Carbolnell,
Few-Body Systems, Suppl. 10, 1999.

\bibitem{4}

"Few-Body Problems in Physics"
Proceedings of the XVIth International IUPAP Conference, Taipei, Taiwan, 
March 6-10, 2000, 
Edited by C.-Y. Cheung, Y.K. Ho, T.-S.H. Lee and S.N. Yang, 
North Holland Elsevier, 2001.

\bibitem{5}
"Few-Body Problems in Physics"
Proceedings of the XVIIth European Conference, Evora, Portugal, 
September 11-16, 2000,
Edited by A. Stadler, A. Arriaga, E. Cravo, A.C. Fonseca, F.M. Nu\~nes, 
M.T. Pe\~na and G. Rupp, 
North Holland Elsevier, 2001. 

\bibitem{6}
"Few-Body Problems in Physics"
Proceedings of the XVIIIth European Conference, Bled, Slovenia, 
September 8-14, 2002, Few-Body Systems, Suppl. 14, 2003. 

\end{thebibliography}
\end{document}